\documentstyle[10pt,aaspp4,psfig]{article}
\begin{document}

\title{Mass and Metallicity of Five X-ray Bright Galaxy Groups}
\author{Una Hwang (1), Richard F. Mushotzky (1), Jack O. Burns (2), Yasushi Fukazawa (3), and Richard A. White (1)} 
\affil{(1) NASA Goddard Space Flight Center, Greenbelt, MD 20771\\ (2)
Department of Physics and Astronomy and Office of Research, University 
of Missouri, Columbia, MO 65211 \\ (3) Department of Physics, University 
of Tokyo, Bunkyo-ku, Tokyo 113, Japan }

\begin{abstract}
We present ASCA X-ray observations of a sample of five groups selected
from a cross-correlation of the ROSAT All-Sky Survey with the White et
al. optical catalog of groups.  These X-ray bright groups
significantly increase the number of known systems with temperatures
between 2 and 3 keV.  They have element abundances of roughly
0.3$\odot$ that are typical of clusters, but their favored ratio of
Si/Fe abundance is lower than the cluster value.  Combining the ASCA
results with ROSAT imaging data, we calculate total masses of a few to
several times $10^{13}\ {\rm M}_\odot$, gas mass fractions of
$\sim$10\%, and baryonic mass fractions of at least 15$-$20\% within a
radius of 0.5 Mpc.  Upper limits for the ratios of gas to galaxy mass
and of the iron mass to galaxy luminosity overlap with the range
observed in rich clusters and extend to lower values, but not to such
low values as seen in much poorer groups.  These results support the
idea that groups, unlike clusters, are subject to the loss of their
primordial and processed gas, and show that this transition occurs at
the mass scale of the 2$-$3 keV groups.  A discussion of ASCA
calibration issues and a comparison of ROSAT and ASCA temperatures are
included in an Appendix.

\end{abstract}

\section{Introduction}

A group with few galaxy members can be readily identified as a
gravitationally bound system rather than a chance superposition of
galaxies on the sky through the detection of X-ray emitting intragroup
gas.  Pointed observations with the Einstein and ROSAT
X-ray Observatories have by now identified several dozen X-ray
emitting groups (e.g., Kriss et al. 1983, Price et al. 1991, Pildis et
al. 1995, Henry et al. 1995, Doe et al. 1995, Mulchaey et al. 1996,
Ponman et al. 1996).
Results using spectrometers of higher resolution and bandwidth on the
ASCA Observatory have now been presented by Fukazawa et al. (1996),
Pedersen et al. (1997), and Davis et al. (1998), among others.  Given
that groups are the most common environment for galaxies, but are
poorly understood compared to their richer cluster cousins, it is
important to enlarge the sample of X-ray emitting groups that have
been well-studied.  The ROSAT All-Sky X-ray Survey (RASS) is expected
to best identify those groups with the highest fluxes, while ROSAT's
limited spectral resolution and bandwidth make data obtained by ASCA
better suited for a spectral study.

In this paper, we present X-ray data from the ASCA Observatory for
five groups of galaxies (N56-395, S34-111, S49-132, N34-175, and
S49-140) selected from a cross-correlation of the RASS with an
optically-selected catalog of groups compiled by White et al. (WBL,
1998; the corresponding WBL catalog numbers are, respectively, WBL518,
WBL025, WBL698, WBL636, and WBL061).  The White et al. catalog was
constructed from the Zwicky catalog of galaxies by an automated
process that mimics the criteria used to create the MKW and AWM
catalogs.  The cross-correlation with the RASS, which provided 98\%
coverage of the X-ray sky at energies from 0.1 to 2.0 keV, was carried
out on a subsample of 68 of the more than 700 groups in the catalog
(Burns et al. 1996).  Our groups are those with the highest X-ray flux
that are also of sufficiently small angular size that the ASCA
instruments (in particular, the SIS in 2-CCD mode) intercept most of
their flux.  These groups should be among the X-ray brightest in the
northern sky.

Our X-ray bright groups have higher temperatures and luminosities and
correspondingly higher masses than the very poor, low temperatures
systems that have been more widely studied.  The low temperature
groups show a wide dispersion in their element abundance and in their
distribution of mass among gas, galaxies, and dark matter.  This has
led to the suggestion that groups are at the threshold to bind their
primordial and processed gas (Fukazawa et al. 1996, Renzini 1997).  We
can test this idea for a somewhat higher mass and temperature range
with our sample of groups.  Throughout this paper, we use $H_0 = 50$
km/s/Mpc.

\section{Data Analysis}

To calculate the mass in various components of the groups, we require
both X-ray spectral and imaging measurements, and optical galaxy
luminosities.  X-ray spectral data from the ASCA Observatory are used
to measure the temperatures and element abundances of the hot X-ray
emitting gas, and to place constraints on the variation of these
quantities with radius.  Imaging data from ROSAT as well as ASCA are
used to characterize the X-ray surface brightness profiles of the hot
gas and to infer the density distributions.  Optical galaxy
luminosities are used to estimate the mass contained in the group
galaxies.  Here we describe the reduction of the X-ray data and review
the optical properties of the groups.

\subsection{Data Reduction}

Data for the groups were obtained with the instruments on the ASCA
Observatory in 1996 and 1998.  The ASCA satellite and its instruments
are described by Tanaka et al. (1994).  Here we briefly note that ASCA
has two Solid-State Imaging Spectrometers (SIS0 and SIS1) and two Gas
Imaging Spectrometers (GIS2 and GIS3), with slightly offset fields of
view.  Each instrument has its own telescope with nested conical foil
mirrors that provide a point-spread function (PSF) with a narrow core
of about $1'$ FWHM, and a half-power diameter of $3'$.  The GIS
broadens the PSF further.  The SIS are square arrays of four CCD chips
that can be used with 1, 2, or all 4 of the chips exposed
simultaneously.  Observations of extended sources are now carried out
primarily in 2-CCD mode because of declining instrument performance in
4-CCD mode.  The field of view of the SIS in 2 CCD mode is $11' \times
22'$.  The GIS field of view is significantly larger ($40'$ useful
diameter), but the GIS provides data of lower spectral and spatial
resolution than does the SIS.  The data were screened and processed
following standard data reduction procedures as outlined in the ASCA
ABC Guide (Day et al., 1997).  Further details on the data processing
can be found in the Appendix.  Table 1 lists the ASCA observations
including the GIS2 exposure time.  We also give the source coordinates
from Ledlow et al. (1996), the nominal Galactic neutral H column
density from Dickey \& Lockman (1990), the group redshift, and the
exposure of available X-ray data from the ROSAT PSPC.

All the groups except N56-395 were also observed by the Position
Sensitive Proportional Counter (PSPC) on the ROSAT Observatory.  The
PSPC provided excellent imaging of 0.2 to 2 keV X-rays combined with
low resolution spectroscopy.  We use software developed by Snowden et
al. (1994) to produce appropriate energy-dependent exposure maps and
to model and subtract the particle background and contamination by
scattered solar X-rays and by the so-called long-term enhancements
(the reader is referred to Snowden et al. for a discussion of these
effects).  The data for S49-132 and S49-140 have also been analyzed by
Doe et al. (1995) as a part of a sample of groups with extended radio
sources, and those for S34-111 (3C31) by Trussoni et al. (1997).  The
ROSAT High Resolution Imager data for S34-111 are presented by
Trussoni et al.; we briefly discuss the 2.1 ks HRI observation of
N34-175 in Section 2.4.

\subsection{ASCA Spectra}

For each group, a circular region on the sky containing most of the
counts was chosen for spectral analysis by examining the GIS image.
These regions generally did not fit completely within the SIS field of
view.  Background spectra were taken from the same detector region on
averaged blank sky fields provided by the ASCA Guest Observer
Facility.  Typically, the individual SIS and GIS spectra had $>$3000
counts after subtraction of the background, but those of S49-140 had
only 900$-$1400 counts, partly because of the exclusive use of FAINT
mode data.  Calibration data from February 1996 and March 1997 were
used to create appropriate instrument response files for each pulse
height spectrum.

Spectra from all four ASCA detectors were fitted simultaneously to the
same spectral model with the relative normalizations allowed to be
fitted freely for each detector.  We use Raymond \& Smith optically
thin thermal models (1977 as updated in XSPECv10.0; hereafter RS) with
the abundances of elements heavier than He taken to be in solar
abundance ratios from Anders \& Grevesse (1989).  The column densities
of SIS0 and SIS1 were allowed to be different from each other, and
were fitted freely, while the column densities for GIS2 and GIS3 were
fixed at the Galactic value (see the Appendix). Figure 1 shows the
global ASCA spectral data for each instrument overlaid with the
best-fit model folded through each instrument response.  Table 2
summarizes the results of these spectral fits.  The errors given in
this and subsequent tables are for 90\% confidence.  The temperatures
are typically 2$-$3 keV, and the gas metallicities are roughly 0.3
$\odot$.  The column densities of SIS0 are listed only for
completeness; they are certainly over-estimated for all but N56-395
(see the Appendix).  The emission measures given are for GIS2; the
GIS3 values are consistent with these to well within 10\%, whereas the
SIS is not useful for obtaining the emission measure because of its
limited field of view.  For N56-395, part of the 12$'$ radius region
used falls outside the useful radius of the GIS; we applied
geometrical correction factors for the emission measure of 1.17 for r
= $4.5'-12'$ and 1.14 for r$<12'$.

We then divided each group into two annular regions to constrain
radial variations in the temperature and abundance.  The width of the
annuli are constrained both by the spatial resolution of the ASCA
mirrors and by the requirement that each spectrum have a sufficient
number of source counts ($\sim 1000$).
Strictly, a spatially resolved analysis of the ASCA spectrum should
account for the broad and energy-dependent point-spread function
(PSF).  The relatively low temperature of our groups reduces the need
for this, as spurious spatial trends arising from neglect of the PSF
are shown to be small for temperatures below about 3 keV
(Takahashi et al. 1995, ASCA News, 3, 34).  As shown in Table 2,
isothermal temperatures and constant element abundances are fully
consistent with this analysis.  Results for S49-140 are not shown in
this table, but are also consistent with isothermal temperatures and
constant abundances.

To place constraints on the relative abundances of the elements, we
fitted the integrated spectrum with a model in which the abundances of
C, N, O, Ne, and Mg are fixed at 0.3 $\odot$, the abundances of Si, S,
Ar and Ca tied to each other in their solar ratios, and Fe tied to Ni.
We thereby obtain a measurement of the relative abundance of
intermediate mass elements represented by Si compared to Fe.  This
ratio is a diagnostic of the enrichment process in the hot gas.  The
results of our fits are shown in Table 3.  All the groups have a ratio
of Si/Fe abundance that is consistent with the solar value.  Contour
plots of the Si abundance {\it vs.} the Fe abundance are shown for
N56-395 and S34-111 in Figure 2.  A ratio of the Si/Fe abundance that
is twice the solar value falls outside the 99\% and 90\% confidence
contours, respectively, for these two groups.

\subsection{Discussion of Spectral Fits}

The robustness of the spectral results can be assessed by comparing
them to the results of fitting each detector individually, since the
simultaneous fitting tends to average out the calibration errors for
individual detectors and give relatively tight errors for the fitted
parameters.  The fitted temperatures typically vary by about 10\%
between the individual detectors with two exceptions.  For S34-111,
the 20\% consistency probably results because, unlike the GIS, the SIS
does not include all the X-ray emitting galaxies in its field of view.
For the outer region of S49-132, a 1$-$2 keV difference between the
SIS and GIS temperatures evidently arises from residual high energy
features in the SIS spectrum.  The SIS temperatures are consistent
with the GIS temperatures when this portion of the spectrum is
ignored, and the joint SIS and GIS fits favor a temperature near the
GIS value.

The single-temperature thermal model that we use is the simplest
possible description of the group X-ray emission consistent with the
ASCA data.  A number of galaxies are detected as X-ray sources, as
seen in Figures 3 and 4.  In addition, some of these galaxies are also
radio sources: N34-175 has a compact radio source associated with the
central galaxy NGC 6338 (e.g., Gregory \& Condon 1991), and may have
nonthermal X-ray emission associated with it as discussed below; the
central galaxy in S34-111, NGC 383, is the radio source 3C31 (see
Trussoni et al. 1997 and references therein); southeast of the center
of S49-132, the galaxy NGC 7503 is an extended radio source with a
head-tail morphology, while the central galaxy in S49-140, NGC 742, is
a head-tail radio source (possibly together with NGC 741; see Doe et
al. 1995 and references therein).

We therefore tried adding a power-law component (presumably emission
from an active galactic nucleus) to the spectral model for each group.
We used a fixed energy index of 1.7 and fixed the column densities at
their values in the fits listed in Table 2.  We found that the fits
for all the groups could be improved in this way, but N34-175 was the
only group for which there was both a significant improvement in the
fit statistic and a significant change in the fitted temperature (see
Tables 2 and 3).  The normalizations of the power-law component in the
two radial annuli are nearly equal, however, whereas the normalization
of the inner annulus should be twice higher if the power-law component
originates from a point-source at the center.  The uncertainties in
the normalizations do not rule out a point-source origin from an AGN,
but it is difficult to rule out the possibility that at least some of
the power-law is a coincidental compensation for residual calibration
deficiencies.  We list results for both a pure thermal spectrum and
for the addition of a power-law for N34-175 throughout.  The 0.5$-$10
keV luminosity of the power-law component is about 2$\times 10^{43}$
ergs/s.  The relative abundance of Si to Fe increases to 1.5 times the
solar value, but is still consistent with the solar value.  The group
S49-140 is also dominated by its central galaxy, but here the 90\%
upper limit on the 0.5$-$10 keV power-law luminosity is only about 6
$\times 10^{41}$ ergs/s.

\subsection{X-ray Images}

Figure 3 shows the Palomar optical survey image of a $18'$ square
region centered on each group with the ASCA GIS2 X-ray contours
overlaid.  Figure 4 shows the same Palomar images overlaid with the
ROSAT PSPC X-ray contours in the approximate energy range 0.5 to 2.0
keV.  There are no PSPC data for N56-395.  The X-ray images in both
figures were smoothed with Gaussian $\sigma=0.6'$ and shifted slightly
to align them with the optical images.  Only S49-132 is markedly
elongated in X-rays, with the elongation following the galaxy
distribution (Doe et al. 1995).  The groups N34-175 and S49-140 are
dominated by their central galaxies.

It is preferable to use the ROSAT PSPC images to characterize the
surface brightness profile of the diffuse X-ray emission because the
ROSAT point-spread-function (PSF) has FWHM=25$''$ at energy 1 keV
while the ASCA mirrors have a broader and more complex spatial
response.  As noted, ROSAT data are available for all the groups
except N56-395.  For S34-111 and S49-132, we mask out the sources from
the 0.5$-$2.0 keV images to form radial surface brightness profiles of
the diffuse emission.  We use 1$'$ bins centered at a position
determined by eye out to a radius of about 50$'$.  The profile of the
diffuse emission is fitted with a King isothermal model for the
surface brightness $\sigma$ as a function of radius $r$:
\[ \sigma (r) = \sigma_0\ \left(1 + \left(\frac{r}{r_c}\right)^2\right)^{-3\beta+\frac{1}{2}} + B_0,\]
where $\beta$ is the slope parameter, $r_c$ the core radius,
$\sigma_o$ the normalization, and $B_0$ a constant background.  The 1
keV ROSAT PSF is convolved with the model, but the results are not
very sensitive to the exact PSF used.  In N34-175 and S49-140, the
central galaxy is particularly bright so it is included in the profile
and modelled as an additional gaussian component.  Other sources are
removed, and 0.5$'$ bins are used.  The radial profiles are shown in
Figure 5, along with the ROSAT HRI profile of N34-175, and the fit
parameters are summarized in Table 4.

Since N56-395 was not observed with the PSPC, we use the ASCA GIS
images to constrain the shape of its surface brightness profile.  We
do not attempt to subtract out the emission from individual sources
because of the extent of the ASCA PSF (the telescope gives a $1'$ core
and half-power diameter of $3'$ and the GIS increases this).  We
generate the appropriate spectrally weighted PSF at the pointing
center by using the GIS pulse-height spectrum as input to a
ray-tracing program.  The fit results are summarized in Table 4 and
Figure 5.  The reliability of the ASCA surface brightness fits for
N56-395 can be tested by comparing such fits for the other groups
against the ROSAT results.  We find good agreement provided that point
sources do not make a large contribution to the flux.  The results for
N56-395 should be fairly reliable since the ROSAT PSPC All-Sky Survey
image of N56-395 does not show strong point sources (Burns et
al. 1996).

From their analysis of ROSAT data, Doe et al. (1995) obtained results
($\beta=0.58\pm0.10$ and $r_c=5.0'\pm0.2'$) in good agreement with
ours for S49-132 .  Their values for S49-140 ($\beta=0.60\pm0.20$ and
$r_c=4.4'\pm0.4'$) are higher than ours, possibly because they simply
exclude the innermost radial bin, whereas we fitted both the diffuse
and the galaxy components in the surface brightness profile.  Trussoni
et al. (1997) use different software and obtain higher values
($\beta=0.61\pm0.06$ and $r_c=7.7'\pm1'$) than we do for S34-111,
but Price et al. (1991) use limited quality Einstein data to obtain
values ($\beta=0.4-0.5$ and $r_c=0.8'-2.5'$) similar to ours.

N56-395 and S34-111 have low fitted values of $\beta \sim$0.4 that are
indicative of relatively flat surface brightness profiles.  The
flatness of S34-111 relative to S49-132, for example, is evident in
Figure 5.  However, the model-fitting carried out here assumes
spherical symmetry, and all the groups show some sign of departure
from spherical symmetry.  The assumption of spherical symmetry for an
elongated gas distribution may result in spuriously low values of
$\beta$ (Makishima et al. 1995), although the most obviously elongated
of our groups (S49-132) actually has the highest fitted value of
$\beta$.  A more complex, two-component profile can also be mimicked
by a single flat profile (Mulchaey \& Zabludoff 1998).

\subsection{Optical Data}

Optically measured velocity dispersions, two-point correlation
coefficients, and luminosities for the group galaxies are summarized
in Table 5.  The velocity dispersions $\sigma$ are the quantity
$S_{BI}$ calculated at 0.75 Mpc from Ledlow et al. (1996) except for
S49-132, for which it is calculated in the conventional way using
velocity and position data in Ledlow et al. for galaxies within a 0.75
Mpc radius circle.  For most of the groups, the velocity dispersions
are based on fewer than 20 or even 10 galaxies, and the true
dispersion may actually lie outside the formal errors.  Zabludoff \&
Mulchaey (1998) show from their deep optical spectroscopic
measurements that velocity dispersions calculated using only the
brightest galaxies tend to significantly underestimate the true
dispersion.  In addition, some of the groups are embedded in larger
systems or are interacting with nearby systems, making the velocity
data difficult to interpret.  For example, S49-132 is embedded in a
larger Zwicky cluster and has an exceptionally high velocity
dispersion.  Its velocity distribution is noted by Ledlow et
al. (1996) and Doe et al. (1995) to be unusually broad and flat, with
three of the central galaxies separated by 1000 km/s in velocity.
N56-395 (MKW 8) is probably interacting with the nearby group MKW 7
(Beers et al. 1995) and its velocity dispersion calculated to 1.5 Mpc
increases substantially to 422$^{+99}_{-53}$ km/s.

The two-point galaxy-galaxy correlation coefficient $B_{gg}$ (see
Andersen \& Owen 1995) is calculated for N galaxies over an angular
scale $\theta$ chosen to optimize signal-to-noise.  It directly
reflects the galaxy richness of the group, and correlates with both
the X-ray luminosity and gas temperature.  S49-132 has the highest
value of B$_{gg}$ among our groups, while N34-175 and S49-140 both
have low B$_{gg}$.

The estimated optical galaxy luminosities are based on B band
magnitudes from the Third Reference Catalog of Bright Galaxies (3RC;
de Vaucouleurs et al. 1991) as accessed through HEASARC and NED.
Wherever available, these are the extrapolated, extinction and
redshift corrected blue magnitudes (B$_{\rm T}^0$).  We include group
galaxies that are located within the radius used for the X-ray
spectral fits, which turns out to be nearly the same as for a uniform
physical radius of 0.5 Mpc.  We do not attempt a correction for faint,
undetected galaxies, and magnitudes are not yet available for many of
the fainter galaxies in the redshift compilation of Ledlow et al., so
our tabulated luminosities are lower limits.

\section{Results}

We now examine physical measurements derived from the data: the X-ray
luminosity and its relationship to the X-ray temperature and to
optically measured dynamical properties; the distribution of mass
between the gas, galaxy, and dark matter components; and the
abundances of the elements.  These are compared to measurements for
other groups (most at lower temperatures than ours) and to rich
clusters to test if our hotter groups provide a transition between
these systems.

\subsection{X-ray Luminosities}

We list in Table 6 the 0.5$-$10 keV X-ray source luminosities measured
within the radii used for the spectral fits.  The luminosities can be
scaled to other radii by using the results of the surface brightness
profile fits.  We also give luminosities at a fixed physical radius of
0.5 Mpc and at a radius representative of the extent of the X-ray
emission.  We define this extent to be the radius where the surface
brightness profile falls to 2\% of its peak value.  The multiplicative
correction factors at these radii are 1.6 at 22$'$ (1 Mpc) for
N56-395, 5.2 at 64$'$ (1.9 Mpc) for S34-111, 1.6 at 20$'$ (1.5 Mpc) for
S49-132, 1.3 at 12$'$ (0.6 Mpc) for N34-175, and 2.4 at 24$'$ (0.8
Mpc) for S49-140.

The X-ray luminosities and temperatures of the groups appear to be
consistent with the well-known correlation in clusters (e.g.,
Mushotzky 1984, Edge \& Stewart 1991, David et al. 1993, Mushotzky \&
Scharf 1997, Fukazawa 1997).  In Figure 6 we show the 0.5$-$10 keV
X-ray luminosity $L_X$ at 0.5 Mpc against the X-ray temperature $kT$
of our groups (square points) compared to a sample of clusters and
groups studied with ASCA by Fukazawa (1997).  The groups fall on the
lower luminosity end of the relation defined by the clusters.  At
temperatures kT $\lesssim$ 1 keV, Ponman et al. (1996) use ROSAT data
for compact groups and find a steeper correlation that they suggest
may arise from the action of galactic winds.  Mulchaey \& Zabludoff
(1998), however, find no evidence for such steepening in their sample
of low temperature (kT $<$ 1 keV) groups.

The X-ray luminosities and various optical measures of the group
richness also appear to follow the same trends seen in richer
clusters.  For example, a strong correlation exists between $L_X$ and
the optical velocity dispersion $\sigma$ in rich clusters (Quintana \&
Melnick 1982, Edge \& Stewart 1991b).  Price et al. (1991) used
Einstein Observatory X-ray data to show that poor clusters and groups
extend the relation found for richer clusters.  The issue has been
complicated since then as Dell'Antonio et al. (1994) and Mahdavi et
al. (1997) suggest that the relation flattens for lower luminosity
groups as the galaxy luminosities begin to dominate over that of the
intragroup gas.  If the galaxy emission is excluded, however, the
cluster $L_X-\sigma$ relation appears to hold for the intragroup gas
(Ponman et al. 1996, Mulchaey \& Zabludoff 1998).  Our groups
generally follow the cluster relation, with the possible exception of
N56-395, for which the velocity dispersion listed in Table 5 is
somewhat low relative to the X-ray temperature and luminosity.  If the
higher dispersion calculated at 1.5 Mpc by Beers et al. (1995) is used
for N56-395, there is better agreement with the average cluster
relation.  Our groups also fall within the range of other clusters and
groups for the relationship between the X-ray luminosity and the
two-point galaxy correlation coefficient B$_{gg}$ shown by Doe et
al. (1995).  The one strong exception is N34-175, for which B$_{gg}$
is markedly low for its X-ray luminosity.  This persists even if the
lower luminosity from the model including a component for the AGN is
adopted.  The relative sparseness of galaxies in this group, which is
reflected in the low value of B$_{gg}$, can be seen in Figures 3 and
4.  N34-175 otherwise shows no anomalies and, as will be shown in the
following section, shows no peculiarities in its mass or baryon
fraction.

With the exceptions noted above, the relations for the X-ray
luminosity to X-ray temperature and to optical richness seen in the
richer clusters appear to hold at the lower mass scales of the kT
$\sim 2-3$ keV groups in our sample.

\subsection{Masses and Mass Ratios}

The mass contained in galaxies, gas, and dark matter in the groups may
be calculated with a few standard assumptions.  The mass of the
galaxies can be estimated from the galaxy luminosities by assuming a
mass-to-light ratio for each galaxy, while the mass of the X-ray
emitting gas and the total gravitational binding mass can be computed
from the X-ray spectral and imaging results by assuming that the gas
is spherically symmetric and in hydrostatic equilibrium.  The
distribution of mass between the various components is then easily
seen by taking ratios of the masses.  These masses and mass ratios are
tabulated in Table 7 both at the radii used for the ASCA fits and at a
fixed physical radius of 0.5 Mpc.

To calculate the mass of a galaxy from its optical B-band luminosity,
an appropriate mass-to-light ratio (M/L) must be determined for it
depending on its galaxy type.  Most of the galaxies are E or S0; for
those with published velocity dispersions and effective (half-light)
radii, we calculate M/L using the scaling in the Fundamental Plane for
elliptical galaxies based on Loewenstein \& White (1998).  Galaxy
velocity dispersions are taken from McElroy (1995), and effective
radii from 3RC.  For galaxies with magnitude information only, we
calculate M/L assuming the mean relation in the Fundamental Plane.  A
few faint (m$_{\rm B} >$ 15) galaxies without morphological
classifications are also treated as ellipticals.  The M/L values are
typically between 4 and 10 in solar units, with a few beyond either
end.  The groups S49-132 and S49-140 each have one relatively bright
galaxy that dominates the mass because of a high M/L ratio arising
from a large effective radius.  If M/L for these galaxies is
calculated from just the luminosity and the mean Fundamental Plane
relation, the total galaxy masses in the lower part of Table 7
decrease from 7.9 to 6.2 $\times 10^{12}$ M$_\odot$ for S49-132, and
from 4.2 to 2.9 $\times 10^{12}$ M$_\odot$ for S49-140.  For the
spiral galaxies, we use the mass-to-light relation given by Persic,
Salucci, \& Stel (1996), which scales with luminosity.  The typical
value of M/L for our spirals is $\sim3$ in solar units.  The galaxy
masses in these groups range from 2 to 8 $\times 10^{12}\ {\rm
M}{_\odot}$.  These are lower limits because optical magnitudes are
not available for many of the known group galaxies and no correction
is attempted for them.

The gas mass is obtained by integrating the hydrogen density over the
appropriate volume and assuming a mean mass per proton of 1.4 amu.
The density in a King isothermal model with slope $\beta$ and core
radius $r_c$ is given by
\[ n (r) = n_{0}\ \left( 1 + \left(\frac{r}{r_c}\right)^2\right)^{-\frac{3}{2}\beta}. \]
The central density $n_{0}$ is determined by equating the spectrally
fitted emission measure with the quantity $\frac{\int{n_e(r) n_H(r)
dV}}{4 \pi d^2}$, where the integral is over the emitting volume,
$d$ is the distance to the source, and $n_e/n_H$ is taken to be 1.23.
The central electron densities are given in Table 4.  The gas masses
at 0.5 Mpc are $(3-8) \times 10^{12}\ {\rm M}_\odot$, as given in
Table 7.

The gravitational binding mass within a radius $r$ depends on the
spatial gradients of the temperature $T$ and the density $\rho$:
\[ M(<r) = \frac{-kT}{G\mu m_p}\ r\ \left( \frac{d\, {\rm ln}\rho}{d\, {\rm ln}r} + \frac{d\, {\rm ln} T}{d\, {\rm ln} r} \right)\, , \]
where k is Boltzmann's constant, G is the gravitational constant,
$m_p$ is the proton mass, and the mean mass per particle, $\mu$, is
taken to be 0.6 amu.  We assume that the gas is isothermal, since this
is consistent with our spatially resolved spectral analysis (Table 2),
and take the gas temperatures from the global spectral results in
Table 2.  Most groups for which the temperature profile has been
measured are indeed consistent with being isothermal (e.g., see
Mulchaey et al. 1996 and references therein, Fukazawa et al.  1996),
with the exception of those whose central regions may harbor cooling
flows (e.g., Fukazawa et al. 1996, 1996b, Ponman \& Bertram 1993).
The total binding masses at 0.5 Mpc for our groups are $(3-8) \times
10^{13}\ {\rm M}_\odot$ (Table 7), higher than those reported for
lower temperature groups by Pildis et al. (1995) and Mulchaey et
al. (1996), but comparable to those reported for six MKW and AWM poor
clusters by Kriss et al. (1983) using Einstein Observatory data.

From the calculated masses, we can compare the distribution of the
total mass between the various components: gas plus galaxy to total
mass (baryon fraction), gas to total mass (gas fraction), and gas to
galaxy mass.  The only significant changes in these quantities between
the ASCA radius and 0.5 Mpc radius are the gas mass fractions for
S34-111 and S49-140.  At the ASCA radius, which correponds to 0.3 Mpc
for both, the gas mass fractions are 30\% and 40\% lower than at 0.5
Mpc.

The baryonic mass fractions of the groups range from 13 to 21\% and
overlap with much of the range for clusters, although some rich
clusters have higher baryon fractions near 30\%.  The gas mass
fractions are between 8 and 14\% and are mostly near 10\%, which is
below the average value for clusters (Allen \& Fabian 1998, White et
al. 1997) but overlaps with the lower end of the cluster range.  Some
other groups, typically those with lower temperature gas, have gas
mass fractions as low as 5\% (e.g., Pedersen et al. 1997, Pildis et
al. 1995, Mulchaey et al. 1996).  At even lower mass scales,
elliptical galaxies have gas mass fractions of less than 1\% (Forman,
Jones, \& Tucker 1985).  There is a progression of increasing gas mass
fraction from elliptical galaxies to poorer groups to clusters,
although it is not clear that this trend continues on cluster scales
(e.g., White et al. 1997).

The upper limits for the ratio of gas mass to galaxy mass in our
groups are between 0.6 and 3.  In clusters, this ratio has been
reported to be as high as 5$-$6 (Arnaud et al. 1992 at 3 Mpc, David et
al. 1990 at 1 Mpc), but others report values of about 1$-$2 (Fukazawa
1997 at 1 Mpc, Edge \& Stewart 1991 at 0.5 Mpc), and trends with mass
on cluster scales are controversial.  Ratios below 1 are reported for
poorer groups (Mulchaey et al. 1996), with many compact groups having
values $<$0.1 (Pildis et al. 1995); elliptical galaxies have ratios as
low as $<$0.01 (Forman et al. 1985).  There is clearly a large-scale
trend of increasing M$_{\rm gas}$/M$_{\rm gal}$ with mass from
ellipticals to poor groups to clusters.  The factor of 5 spread we
find for our groups spans the range for rich clusters and goes
significantly below it, showing that the deviations from the cluster
values take place at this 2$-$3 keV mass scale.  The values are
significantly higher than the typical values for poorer groups,
including the compact groups.  The galaxy masses for the groups are
lower limits, but it is unlikely that they are in error by
substantially more than a factor of 2, and corrections to them will
move the gas to galaxy mass ratios farther from the cluster values.
Results in the literature are reported rather nonuniformly and there
are few observations that extend to the virial radii of groups and
clusters, but the large-scale trends noted here seem to be clearly
established.

Errors in the gas mass due to the formal uncertainties in $\beta$ and
$r_c$ and in $kT$ are typically smaller than a few percent, while the
formal errors in the total mass are about 10\% or less.  N56-395 has
the largest uncertainties in $\beta$ and $r_c$, and the error in its
ratio of gas mass to total mass is about 13\%.  If we use the ROSAT
results of Trussoni et al. (1997) for S34-111, with $\beta$ and $r_c$
about twice the values we obtain, the masses are still within about
10\% of the values in Table 7.  The systematic uncertainties in the
mass calculations are likely to be comparable to or higher than the
formal uncertainties.  If the assumption of hydrostatic equilibrium
does not hold, the mass calculations may be in error by some 30\%
(Evrard, Metzler, \& Navarro 1996, Schindler 1996), and the neglect of
an existing temperature gradient will contribute additional errors.

A simple check on our calculations for the total mass is to compare
them to the virial mass derived from the scaling of mass and
temperature in the simulations of Evrard, Metzler, \& Navarro (1996).
They provide analytical formulae depending only on the temperature of
the X-ray emitting gas both for the virial radius, defined as
enclosing a mass density contrast of 500 times the critical density,
and for the corresponding mass.  For 2 and 3 keV gas, these radii are
2.2 and 2.7 Mpc, and the masses are 2$\times 10^{14}\ {\rm M}_\odot$
and 3.65$\times 10^{14}\ {\rm M}_\odot$, respectively.  Extrapolating
our results to these radii, we derive masses that are in excellent
agreement with the analytical predictions, with the exception of
N34-175.  Its extrapolated mass of 3$\times 10^{14}\ {\rm M}_\odot$
for a temperature near 2 keV is somewhat high.  Using the lower 1.6
keV temperature from fitting an power-law component, the extrapolated
mass of 2$\times 10^{14}\ {\rm M}_\odot$ is still high relative to the
analytical mass of 1.4$\times 10^{14}\ {\rm M}_\odot$.

The gas mass obtained by Doe et al. (1995) for S49-132 is difficult to
reconcile with ours.  They process the ROSAT imaging data using
methods similar to ours, and obtain $\beta$ and $r_c$ values in good
agreement, but calculate the mass based on a deprojection of the ROSAT
data.  Much of the difference between our calculated gas masses is
accounted for by their calculated central electron density being a
factor of three higher than ours.  The higher density would be
required for the low ROSAT temperature of 2 keV they use compared to
the ASCA temperature of 3 keV.  At 0.5 Mpc, their X-ray luminosity is
nearly an order of magnitude lower than ours, and their gas mass
fraction of 45\% is much higher than our value.  For S49-140, our
results are in reasonable agreement once the total mass is corrected
for the ASCA temperature of 1.6 keV compared to the temperature 1.0
keV used by Doe et al., notwithstanding our discrepancies in $\beta$
and $r_c$.

\subsection{Element Abundances}

Our groups have average element abundances of about $0.3 \odot$, which
is the remarkably homogeneous value observed in the rich clusters
(Allen \& Fabian, 1998, Mushotzky et al. 1996 and references therein).
Although some other groups have cluster-like abundances (e.g.,
Fukazawa et al. 1996), others have substantially lower element
abundances (e.g., NGC 4261, Davis et al. 1995; NGC 2300, Davis et
al. 1996).  Low abundance groups have now been confirmed (NGC 2300,
Sakima et al. 1995) and newly identified by ASCA (e.g., NGC 3259,
Pedersen et al. 1997; also see Davis et al. 1998).  Fukazawa (1997)
and Renzini (1997) point out a pattern in the X-ray temperature and Fe
abundance of groups and clusters: at temperatures above about 1 keV,
the Fe abundance shows little variance from about 0.3 times solar,
while at lower temperatures, there is a large spread in abundances,
with a tendency toward very low values\footnote{It is possible that
some of this spread is related to uncertainties in the atomic physics
of the Fe L shell, which dominates the abundance and temperature
determination at low temperatures (see references above).}.  Our
groups fit into the pattern noted by Fukazawa and Renzini in that a
0.3 $\odot$ metallicity is expected for temperatures of 2$-$3 keV.

The Si/Fe abundance ratio in our groups favors the solar value,
although a ratio of twice the solar value cannot in general be ruled
out with more than 90\% confidence.  In other groups for which the
Si/Fe abundance ratio has been determined so far, this ratio is
generally about solar (Fukazawa et al. 1996, Davis et al. 1998).  This
is similar to the value determined for elliptical galaxies (Matsumoto
et al. 1997), and lower than the value of twice solar characteristic
of rich clusters (Mushotzky et al. 1996).

The quantity M$_{\rm Fe}$/L$_{\rm B}$ is even more directly related to
the stellar Fe yield than the abundances themselves (Arnaud et
al. 1992).  As shown in Table 7, the upper limits for our groups are
between 0.005 and 0.015 in solar units when evaluated at a radius of
0.5 Mpc.  Since this quantity is directly proportional to the Fe
abundance, its errors are dominated by the uncertainties in the Fe
abundance measurement.  These errors range from less than 20\% up to
45\% in our groups (Table 2).  The typical range of M$_{\rm
Fe}$/L$_{\rm B}$ in clusters is 0.01$-$0.02, measured at a radius of 3
Mpc, roughly the virial radius for clusters (Renzini 1997).
Comparisons at different radii are a little risky, but the values for
our groups are evidently at or below the typical range for clusters.
In our sample, there appears to be a significant spread in this
quantity ranging from the cluster value to below.  The mass scale
corresponding to temperatures of 2 to 3 keV is evidently that at which
the M$_{\rm Fe}$/L$_{\rm B}$ ratio begins to deviate from the cluster
value.  In the larger sample of Davis et al. (1998), two groups with
temperatures $\sim$ 2 keV have even lower M$_{\rm Fe}$/L$_{\rm B} <$
0.001.  The pattern of increasing M$_{\rm Fe}$/L$_{\rm B}$ in poor
groups to rich clusters echoes that seen in the lower gas mass
fractions and gas to galaxy mass ratios of groups compared to
clusters.

\section{Conclusions}

Our study of a sample of five RASS-selected, X-ray bright groups shows
that these objects have temperatures between 2 and 3 keV at the lower
boundary of the temperature range for clusters.  This is a temperature
range for which relatively few objects have been well-studied thus far
compared to those above and below it.  Insofar as we are able to test
the spatial temperature structure by comparing the spectra in two
annular rings, it is consistent with being isothermal out to about
10$'$.  The temperatures and bolometric luminosities agree well with
the relation established for richer clusters, reinforcing the
conclusions of other authors that the bulk X-ray properties of groups
are an extension of those of richer clusters (Price et al. 1991, Doe
et al. 1995, Burns et al. 1996, Mulchaey \& Zabludoff 1998).  We also
find that both the optically determined galaxy velocity dispersion and
the galaxy richness parameter $B_{gg}$ compared to the X-ray
luminosity generally agree well with the relations established by
other clusters and groups.  The dynamically complex group N56-395 has
a relatively low velocity dispersion at 0.75 Mpc radius, whereas
N34-175 has a low $B_{gg}$ for its X-ray luminosity.

The spectral and imaging data together allow estimates of the gas mass
and the total gravitating mass in these systems.  At a radius of 0.5
Mpc, the gas masses are a few to several times $10^{12}\ {\rm
M}_\odot$, and the total masses are about a factor of ten higher,
giving gas mass fractions near 10\% for this sample.  The ratio of
mass in gas relative to that galaxies spans a factor of several and is
no higher than about 0.6$-$3.  These values overlap the range seen in
clusters but are higher than those seen in lower temperature groups.
Whereas poorer groups tend to have widely varying and often rather
low abundances, our groups have element abundances typical of
clusters, at roughly 0.3 $\odot$.  The ratio of Si/Fe abundance,
however, favors at 90\% confidence a value near the solar ratio that
is more characteristic of elliptical galaxies and poorer groups than
of rich clusters.  The ratio of Fe mass in the gas to the optical
luminosity spans a factor of three range from 0.005 to 0.015.  These
values also overlap those for the hotter clusters and are higher than
the widely dispersed values reported for lower temperature groups.
The 2$-$3 keV groups appear to be at the mass scale at which the
transitions occur from the cluster values to the lower, widely
dispersed in values observed in the poorer, lower temperature groups.

The uniformity of cluster element abundances and M$_{\rm Fe}$/L$_{\rm
B}$ indicates that clusters had very similar star formation histories
involving little or no exchange of baryons with their surroundings
(Renzini 1997).  Groups are a loosely defined class of objects,
ranging from isolated pairs to aggregates comparable to the poorest
clusters, and this diversity is reflected in the large dispersion in
their properties.  The wide range in observed element abundance,
M$_{\rm Fe}$/L$_{\rm B}$, and in the distribution of mass between gas,
galaxies, and dark matter indicates a varied formation and chemical
enrichment history for groups.  Groups, which have lower gravitational
potentials than the more massive clusters, must have lost gas to their
surroundings due to galactic winds or the supernova heating that
followed star formation (Fukazawa et al. 1996, Renzini 1997). Such an
exchange is consistent with the lower M$_{\rm Fe}$/L$_{\rm B}$ ratios,
lower gas mass fractions, and lower gas to galaxy mass ratios measured
for groups.  The low abundances observed in some poor groups may
indicate that these groups later accreted pristine gas that diluted
the abundances (Renzini 1997), or perhaps simply that these groups
have lost more of their enriched gas (e.g., Davis et al. 1998).

The X-ray bright groups selected from the RASS-optical catalog
correlation are at the upper end of the mass scale for groups with
masses, temperatures, luminosities, and abundances comparable to the
poorest of the Abell clusters.  In properties such as M$_{\rm
Fe}$/L$_{\rm B}$ and the balance of mass between gas and galaxies,
they begin to deviate from the cluster norms.  Our results reinforce
the suggestion that groups are at the threshold to confine their
primordial and processed gas for a less widely studied mass range
intermediate between the poorest groups and the rich clusters.

Groups are finally receiving the attention they deserve based on their
preponderance in the Universe.  X-ray studies are crucial, and
improved X-ray observations with higher spatial resolution are highly
desirable, both to increase the sample of studied groups and also to
separate out contaminating emission from individual galaxies.  The
X-ray emission from galaxies is relatively much more important in
groups and in poor clusters and can make up a substantial fraction of
the flux.  In the case of N34-175, emission from the AGN increases the
uncertainty in the temperature measurement significantly.  Such
improved observations are forthcoming with AXAF and XMM.

\acknowledgments

We are grateful to Andy Ptak for providing software for simulating the
ASCA point spread function, fitting image profiles, and plotting image
contours.  We thank Stephen Doe, Kurt Roettiger, and especially Mike
Loewenstein for helpful scientific discussions, Mark Bliton for
calculating B$_{\rm gg}$ for N34-175, Koji Mukai and Tahir Yaqoob for
discussion and help with ASCA calibration and data processing issues,
and Steve Snowden for discussion of ROSAT processing issues.  We also
thank the referee for his comments.  This work made use of the
HEASARC, NED, and SkyView databases.  Partial support was provided by
NASA grant NAG 5-3410 to J.O.B. and R.A.W.

\newpage
\centerline{\bf APPENDIX: \ CALIBRATION ISSUES }

Radiation damage to the ASCA CCD detectors is causing an increasingly
marked decline in instrument performance that is most evident in 2-
and 4-CCD mode.  The most notable effect is a reduced efficiency in
the SIS detectors at energies below about 1.2 keV that is mimicked by
a spuriously high column density when the data are fitted with a
spectral model.  The effect is worse for SIS1 than for SIS0, and
current calibration corrections do not completely compensate for it.
Furthermore, some important calibration corrections (residual dark
current, or RDD, and dark frame error, or DFE) can be applied only to
data obtained in FAINT mode\footnote{Discussion of the RDD, DFE, and
the characteristics of FAINT and BRIGHT mode data can be found in the
ASCA ABC Guide (Day et al., 1997).}.  Along with the change in column
density, sometimes there is a significant change in the fitted
temperature between BRIGHT and FAINT mode.  For this reason, we
present only the FAINT mode results as being more reliable.
Furthermore, different algorithms for implementing the FAINT mode
corrections give somewhat different values for the fitted column
density.  We used the method that gave lower $N_H$, since the
temperatures were consistent.  We stress that the column densities
obtained from fits to these data should not be considered reliable
measurements.

Because of the higher telemetry requirements for obtaining data in
FAINT mode over BRIGHT mode, most of our observations---scheduled
before the full effect of declining SIS performance was known---were
carried out in a combination of FAINT and BRIGHT mode.  Limiting the
spectral analysis to FAINT mode data decreases the SIS exposure time
for S49-132, N34-175, and S49-140 to, respectively, 75\%, 50\%, and
45\% of those listed in Table 1 for GIS2.  For S34-111, the FAINT and
BRIGHT mode data gave consistent results so we use BRIGHT mode to
improve the statistics.  

ROSAT PSPC data are available for four of our five groups.  We
examined the spectral data to determine the column densities.  We
extracted the PSPC spectra from the same region on the sky as was used
for the ASCA spectra and retained all point sources (since we did not
subtract point sources from the ASCA spectra).  The column densities
are in excellent agreement with the Galactic values, but the
temperatures are all lower than the ASCA temperatures by about 30\%.

Figure 7 summarizes all the available data.  It plots against the date
of the ASCA observation measured in months since the ASCA launch
(February 1993): (1) the column densities obtained by spectral fits to
our SIS0 and SIS1 data using BRIGHT mode, (2) from fits to the ROSAT
PSPC data, and (3) the Galactic column density from Table 1.  It is
clearly seen that the fitted SIS column densities exceed the ROSAT and
Galactic values, increase with time, and are higher for SIS1 than for
SIS0.  We have verified that this is not caused by a high CCD
temperature or by inappropriate background subtraction.

As noted above, the ROSAT temperatures for the groups are
systematically lower than the ASCA temperatures.  In Figure 8, we
compare the temperature results for our groups with a representative
(not comprehensive) sample of ROSAT and ASCA temperatures for
elliptical galaxies and clusters taken from the literature.  The 30\%
offset between ROSAT and ASCA temperatures for our groups is typical
for clusters with temperatures above about 2 keV.  There does not
appear to be any systematic offset at lower temperatures, and there is
a hint that the agreement may improve at temperatures above about 6
keV.  A number of effects may contribute to the discrepancies.  It may
simply reflect the limitations of a low-resolution, small-bandwidth
spectrometer for accurately measuring gas temperatures above about 2
keV.  Another possibility is that the gas has a wide range of
temperatures.  ROSAT's lower energy bandwidth would make it more
sensitive to the lower temperature gas, while ASCA's broader energy
bandwidth would make it measure a higher average temperature.


\newpage

\begin{deluxetable}{llllccll}
\tablecaption{X-ray Observations}
\tablewidth{0pt}
\tablehead{
\colhead{Source} &\colhead{Observation} & \colhead{Exposure} & \colhead{RA} & \colhead{DEC} &\colhead{Gal N$_{\rm H}$\tablenotemark{a}} & \colhead{Redshift\tablenotemark{b}} & \colhead {Exposure}\nl
\colhead{} & \colhead{Date} & \colhead{GIS2 (ks)} & \colhead{(J2000)} & \colhead{(J2000)} & \colhead{(10$^{21}$ cm$^{-2}$)} & \colhead{} & \colhead{RPSPC (ks)} \nl
}
\startdata
N56-395 &96 Feb 9$-$10  &20.2 & 14 40 43.2 & 03 27 12 &0.28&0.0272 & ...  \nl
S34-111 &96 Aug 9       &23.5 & 01 07 27.7 & 32 23 59 &0.54&0.0173 & 24.9 \nl
S49-132 &96 Dec 11      &17.5 & 23 10 31.8 & 07 34 18 &0.49&0.0419 & 12.6 \nl
N34-175 &98 Feb 18$-$19 &26.1 & 17 15 21.4 & 57 22 43 &0.26&0.0283 & 3.8  \nl
S49-140 &98 Aug $1-2$   &32.8 & 01 56 22.9 & 05 37 37 &0.44&0.0179 & 12.9 \nl
\enddata
\tablenotetext{a}{Dickey \& Lockman 1990, ARAA, 28, 215}
\tablenotetext{b}{Ledlow et al. 1996, S49-132 from data therein}
\end{deluxetable}

\begin{deluxetable}{llllllc}
\tablecaption{X-ray Spectral Fits I\tablenotemark{a}}
\tablewidth{0pt}
\tablehead{
\colhead{Source} & \colhead{Radius} & \colhead{$\chi^2$} & \colhead{N$_{\rm H}$, SIS0\tablenotemark{b}} & \colhead{kT} & \colhead{Abundance} & \colhead{EM, GIS2} \nl
\colhead{} & \colhead{(arcmin)} & \colhead{} & \colhead{(10$^{21}$ cm$^{-2}$)} & \colhead{(keV)} & \colhead{(rel $\odot$)} & \colhead{($10^{11}\ \rm{cm}^{-5}$)} \nl
}
\startdata
N56-395 & $r<12$      & 483.1, 0.93 & 0.4 (0.3$-$0.6) & 3.11 (2.99$-$3.24) & 0.34 (0.27$-$0.42) & 18 \nl
        & $r<4.5$     & 327.0, 1.14 & 0.3 ($<$0.5) & 3.26 (3.09$-$3.43) & 0.38 (0.28$-$0.48) & 6.3 \nl
        & $4.5<r<12$  & 342.9, 1.00 & 0.6 (0.3$-$1.0) & 3.05 (2.86$-$3.24) & 0.31 (0.20$-$0.43) & 11 \nl
 & & & & & & \nl
S34-111 &$r<10.5$     & 355.9, 1.00 & 0.4 (0.1$-$0.8) & 2.10 (1.99$-$2.22) & 0.37 (0.26$-$0.50) & 7.7 \nl
        & $r<4.5$     & 157.8, 1.07 & 0.6 (0.2$-$1.0) & 2.25 (2.10$-$2.45) & 0.33 (0.18$-$0.51) & 2.5 \nl
        & $4.5<r<10.5$& 253.6, 1.04 & 0.4 ($<$0.9) & 2.07 (1.90$-$2.22) & 0.58 (0.38$-$0.85) & 4.3 \nl
 & & & & & & \nl
S49-132  & $r<9$      & 393.9, 1.05 & 1.0 (0.7$-$1.3) & 3.37 (3.20$-$3.54) & 0.23 (0.14$-$0.33) & 10 \nl
 & $r<4$      & 178.0, 0.95 & 1.0 (0.6$-$1.5) & 3.10 (2.88$-$3.33) & 0.27 (0.14$-$0.41) & 4.1 \nl
 & $4<r<9$    & 270.7, 1.08 & 1.0 (0.6$-$1.5) & 3.53 (3.27$-$3.83) & 0.18 (0.04$-$0.32) & 6.4 \nl
 & & & & & & \nl
N34-175  & $r<7.5$    & 709.6, 1.43 & 1.3 (1.1$-$1.6)   & 2.25 (2.19$-$2.31) & 0.42 (0.35$-$0.49) & 14 \nl
 & $r<3$      & 340.0, 1.28 & 1.2 (0.9$-$1.4) & 2.26 (2.18$-$2.42) & 0.54 (0.43$-$0.67) & 6.1 \nl
 & $3<r<7.5$  & 430.6, 1.36 & 1.4 (1.0$-$1.7) & 2.26 (2.17$-$2.42) & 0.35 (0.26$-$0.46) & 7.9 \nl
 & & & & & & \nl
N34-175 (with PL) & $r<7.5$ & 631.9, 1.27 & 1.3 (fixed) & 1.63 (1.53$-$1.71) & 0.35 (0.28$-$0.43) & 11 \nl
 & & & & & & \nl
S49-140  & $r<8.5$    & 259.2, 1.07 & 0.1 ($< 0.7$)   & 1.56 (1.41$-$1.68) & 0.32 (0.21$-$0.47) & 3.2 \nl
\enddata
\tablenotetext{a}{Errors in this and subsequent tables are for 90\% confidence ($\Delta\chi^2=2.7$).}
\tablenotetext{b}{See the Appendix.}
\end{deluxetable}

\begin{deluxetable}{lllll}
\tablecaption{X-ray Spectral Fits II}
\tablewidth{0pt}
\tablehead{
\colhead{Source} & \colhead{$\chi^2$} & \colhead{kT} & \colhead{Si} & \colhead{Fe} \nl
\colhead{} & \colhead{} & \colhead{(keV)} & \colhead{(rel $\odot$)} & \colhead{(rel $\odot$)} \nl
}
\startdata
N56-395 & 483.1, 0.93 & 3.12 (2.99$-$3.25) & 0.30 (0.10$-$0.50) & 0.34 (0.27$-$0.41) \nl
S34-111 & 353.1, 1.00 & 2.08 (1.96$-$2.20) & 0.49 (0.30$-$0.71) & 0.39 (0.29$-$0.51) \nl
S49-132 & 393.4, 1.05 & 3.38 (3.21$-$3.56) & 0.16 ($<$0.46)     & 0.22 (0.13$-$0.31) \nl
N34-175 & 643.8, 1.40 & 2.26 (2.19$-$2.45) & 0.27 (0.14$-$0.40) & 0.41 (0.34$-$0.48) \nl
N34-175 (with PL) & 577.3, 1.25 & 1.52 (1.41$-$1.64) & 0.50 (0.39$-$0.63) & 0.34 (0.27$-$0.42) \nl
S49-140 & 257.4, 1.07 & 1.41 (1.33$-$1.64) & 0.45 (0.28$-$0.65) & 0.27 (0.18$-$0.34) \nl
\enddata
\end{deluxetable}

\begin{deluxetable}{llll}
\tablecaption{X-ray Spatial Properties}
\tablewidth{0pt}
\tablehead{
\colhead{Source} & \colhead{$\beta$} & \colhead{r$_{\rm core}$} & \colhead{n$_{\rm e,0}$}\nl
\colhead{} & \colhead{} & \colhead{(arcmin)} & \colhead{$10^{-3}\ {\rm cm}^{-3}$}\nl
}
\startdata
N56-395\tablenotemark{a} & 0.40 $^{+0.03}_{-0.04}$& 1.3 $^{+0.5}_{-0.4}$ & 3.7$^{+0.4}_{-1.1}$\nl
S34-111 & 0.38 $\pm$ 0.02 & 3.0 $\pm 0.5$ & 1.3$\pm 0.1$\nl
S49-132 & 0.63 $^{+0.05}_{-0.04}$ & 5.0 $^{+0.6}_{-0.5}$ & 1.1$\pm$0.1 \nl
N34-175 & 0.60 $^{+0.05}_{-0.04}$& 2.8 $\pm$ 0.5 & 3.1$\pm 0.4$\nl
S49-140 & 0.39 $^{+0.05}_{-0.03}$& 1.3 $^{+0.9}_{-0.7}$ & 2.2$^{+0.8}_{-1.2}$\nl
\enddata
\tablenotetext{a}{Using the ASCA GIS}
\end{deluxetable}

\begin{deluxetable}{lllllcc}
\tablecaption{Optical Properties}
\tablewidth{0pt}
\tablehead{
\colhead{Source} & \colhead{Other Name} & \colhead{$\sigma$\tablenotemark{a}} & \colhead{B$_{\rm gg}$} & N & $\theta$ & \colhead{L$_{\rm B}$\tablenotemark{c}} \nl
 & & (km/s) & & & (arcmin) &  \colhead{($10^{11}\ {\rm L}_\odot$)} \nl
}
\startdata
N56-395 & WBL518, MKW 8 & 329$^{+110}_{-97} (15) $ & 101 $\pm$ 33 & 11 & 27.9 &$>$3.1 (5) \nl
S34-111 & WBL025        & 466$^{+55}_{-41}$ (29)  & 100 $\pm$ 33 & 10 & 41.6 &$>$4.3 (12) \nl
S49-132 & WBL698        & 877 (21)              & 156 $\pm$ 39 & 18 & 19.3 &$>$7.3 (6) \nl
N34-175 & WBL636        & 589$^{+440}_{-31}$ (7)  & 49 $\pm$ 22  & 6  & 25.8 &$>$4.0 (5) \nl
S49-140 & WBL061        & 205$^{+59}_{-21}$  (10) & 58 $\pm$ 36\tablenotemark{b}  & 7  & 36.4 &$>$2.6 (7) \nl
\enddata
\tablenotetext{a}{S$_{\rm BI}$ at 0.75 Mpc, Ledlow et al. (1996); S49-132 from data in Ledlow et al. (1996).  In parentheses is the number of galaxy velocities used in the calculation.}
\tablenotetext{b}{Andersen \& Owen (1995)}
\tablenotetext{c}{L$_{\rm B}$ at 0.5 Mpc radius.  In parentheses is the number of galaxy magnitudes used.}
\end{deluxetable}
 

\begin{deluxetable}{lccccccc}
\tablecaption{0.5$-$10 keV X-ray Luminosities\tablenotemark{a}}
\tablewidth{0pt}
\tablehead{
\colhead{Source} & \colhead{R$_{\rm X}$} & & \colhead{L$_{\rm x, R_X}$} & \colhead{R$_{\rm 0.5 h^{-1} Mpc}$} & \colhead{L$_{\rm x, 0.5 Mpc}$} & \colhead{Extent} & \colhead{L$_{\rm x, R_E}$}  \nl
\colhead{} & \colhead{(arcmin)} & \colhead{($h^{-1}$ Mpc)} & \colhead{($h^{-2} 10^{43}$ ergs/s)} & (arcmin) & \colhead{($h^{-2} 10^{43}$ ergs/s)} & \colhead{($h^{-1}$ Mpc)} & \colhead{($h^{-2} 10^{43}$ ergs/s)}  \nl
}
\startdata 
N56-395 &12  &0.57 & 6.4 & 10.5 & 5.8 & 1.0 & 10. \nl
S34-111 &10.5&0.32 & 1.0 & 16.4 & 1.5 & 1.9 & 5.2 \nl
S49-132 &9   &0.66 & 8.9 & 6.8  & 6.9 & 1.5 & 14. \nl 
N34-175 &7.5 &0.37 & 5.0 & 10.1 & 5.9 & 0.6 & 6.3 \nl
(with PL)&...&...  & 3.3 & ...  & 3.9 & ... & 4.2 \nl
S49-140 &8.5 &0.27 & 0.4 & 16.0 & 0.7 & 0.8 & 0.9 \nl
\enddata
\tablenotetext{a}{H$_{\rm 0}$ = 50 km/s/Mpc}
\end{deluxetable}

\begin{deluxetable}{lcccccccc}
\tablecaption{Masses\tablenotemark{a}}
\tablewidth{0pt}
\tablehead{
\colhead{Source} & \colhead{M$_{\rm gas}$} & \colhead{M$_{\rm opt}$} & \colhead{M$_{\rm Fe}$/L$_{\rm B}$}& \colhead{${\rm M_{gas}/M_{gal}}$} & \colhead{M$_{\rm bind}$} & \colhead{$f_{\rm gas}$} & \colhead{$f_{\rm baryon}$} \nl
\colhead{}  & \colhead{($h^{-5/2} 10^{12}\ {\rm M}_\odot$)} & \colhead{($h^{-1} 10^{12}\ {\rm M}_\odot$)} & \colhead{(M$_\odot$/L$_\odot$)} & & \colhead{($h^{-1} 10^{12}\ {\rm M}_\odot$)} & \colhead{} & \colhead{} \nl
}
\startdata 
To R$_{\rm X}$:  & & & & & & & \nl
N56-395  & 8.9 &$>$2.2 & 0.018 &$<$4.0 & 78 & 0.11 &$>$0.14 \nl
S34-111  & 1.6 &$>$2.9 & 0.003 &$<$0.6 & 26 & 0.06 &$>$0.17 \nl
S49-132  & 15  &$>$7.9 & 0.009 &$<$1.9 & 120& 0.12 &$>$0.19 \nl
N34-175  & 4.7 &$>$2.9 & 0.009 &$<$1.6 & 49 & 0.10 &$>$0.16 \nl
(with PL)& 4.2 & ...   & 0.007 &$<$1.4 & 35 & 0.12 &$>$0.20 \nl
S49-140  & 0.8 &$>$3.8 & 0.002 &$<$0.2 & 18 & 0.05 &$>$0.26 \nl
To 0.5 Mpc:  & & & & & & & \nl
N56-395  & 6.9 &$>$2.2 & 0.014 &$<$3.1 & 68 & 0.10 &$>$0.13 \nl
S34-111  & 3.9 &$>$2.9 & 0.006 &$<$1.3 & 42 & 0.09 &$>$0.16 \nl
S49-132  & 8.2 &$>$7.9 & 0.005 &$<$1.0 & 76 & 0.11 &$>$0.21 \nl
N34-175  & 7.7 &$>$2.9 & 0.015 &$<$2.6 & 70 & 0.11 &$>$0.15 \nl
(with PL)& 6.8 & ...   & 0.011 &$<$2.3 & 50 & 0.14 &$>$0.19 \nl
S49-140  & 2.7 &$>$4.2 & 0.006 &$<$0.6 & 34 & 0.08 &$>$0.20 \nl
\enddata
\tablenotetext{a}{H$_{\rm 0}$ = 50 km/s/Mpc}
\end{deluxetable}

\begin{figure}
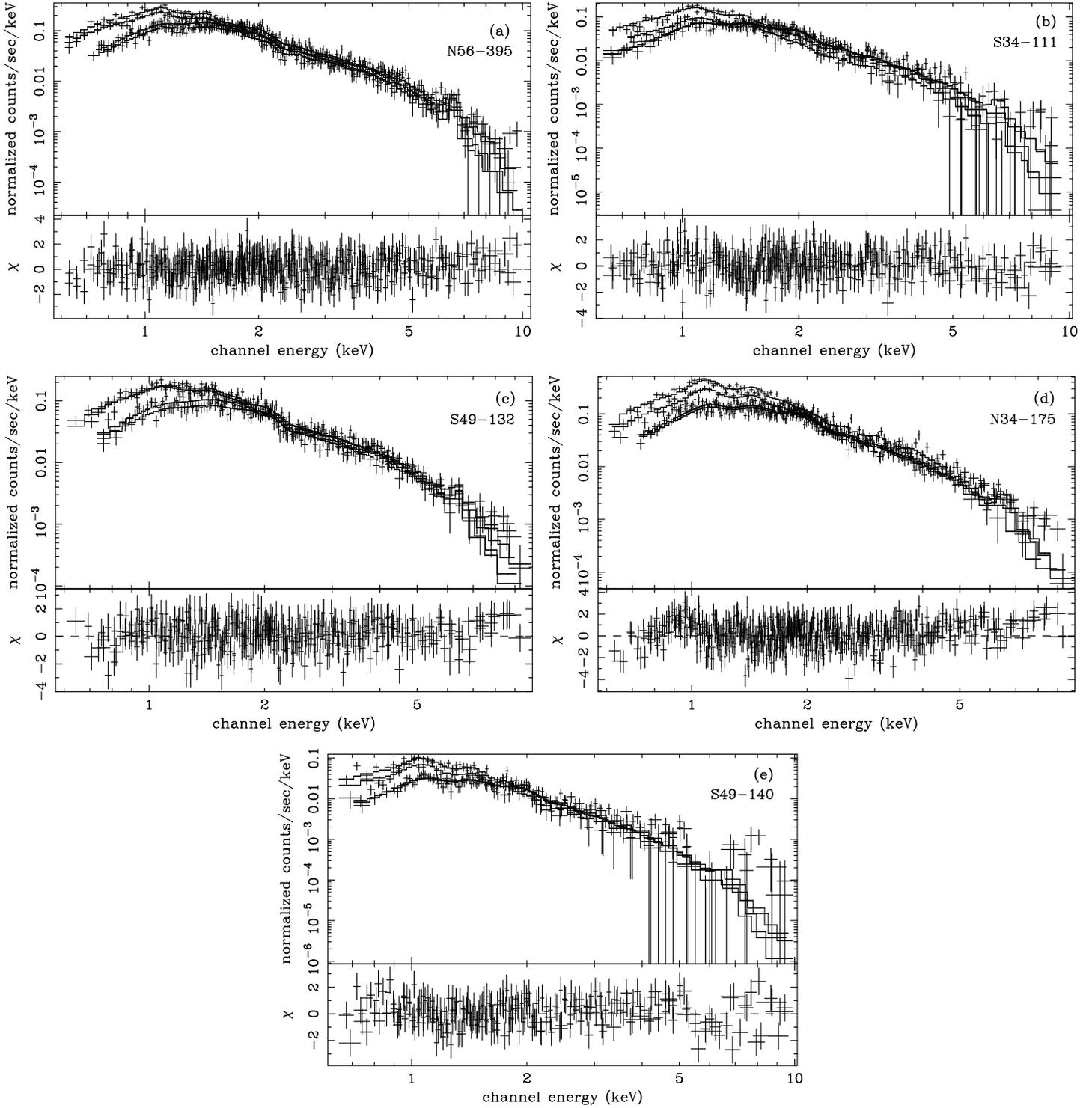

\centerline{
\hfill 
\psfig{figure=f1a.ps,height=2.5truein,angle=-90} \
\psfig{figure=f1b.ps,height=2.5truein,angle=-90} 
\hfill}
\vskip 0.1in
\centerline{
\psfig{figure=f1c.ps,height=2.5truein,angle=-90} \
\psfig{figure=f1d.ps,height=2.5truein,angle=-90} 
\hfill}
\vskip 0.1in
\centerline{
\psfig{figure=f1e.ps,height=2.5truein,angle=-90} 
}\caption{ASCA spectral data (SIS0, SIS1, GIS2 and GIS3) overlaid
with the best-fit single-component, single abundance, RS spectral
model folded through each instrument response. 
The lower panels show the residuals.}
\end{figure}

\begin{figure}
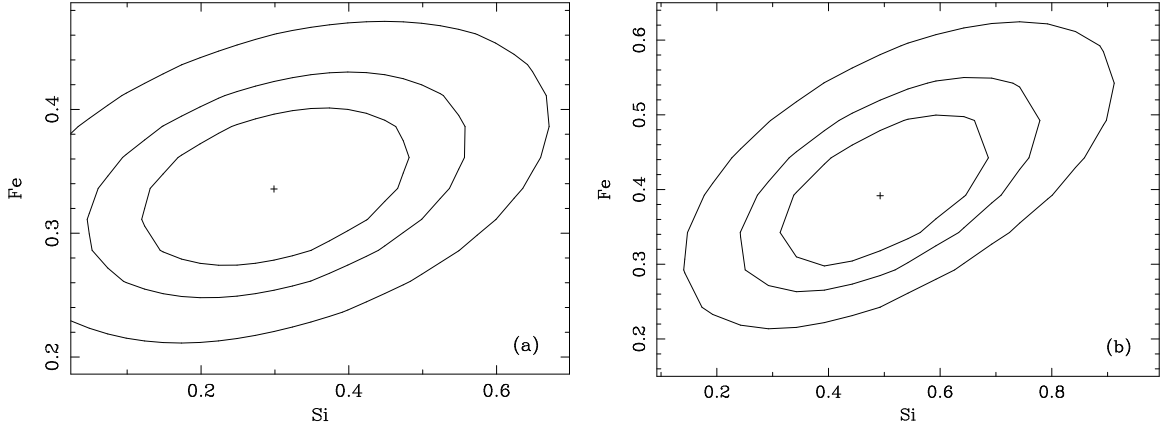

\centerline{
\hfill 
\psfig{figure=f2a.ps,height=2.2truein,angle=-90} \ \
\psfig{figure=f2b.ps,height=2.2truein,angle=-90} 
\hfill
}\caption{Two-dimensional $\Delta\chi^2$ contours for the Si and Fe
abundances in (a) N56-395 and (b) S34-111.  The contours correspond to
$\Delta\chi^2$ = 2.3, 4.61, 9.21 (68\%, 90\%, and 99\% confidence for
two parameters).}
\end{figure}

\begin{figure}
\centerline{
\hfill 
\psfig{figure=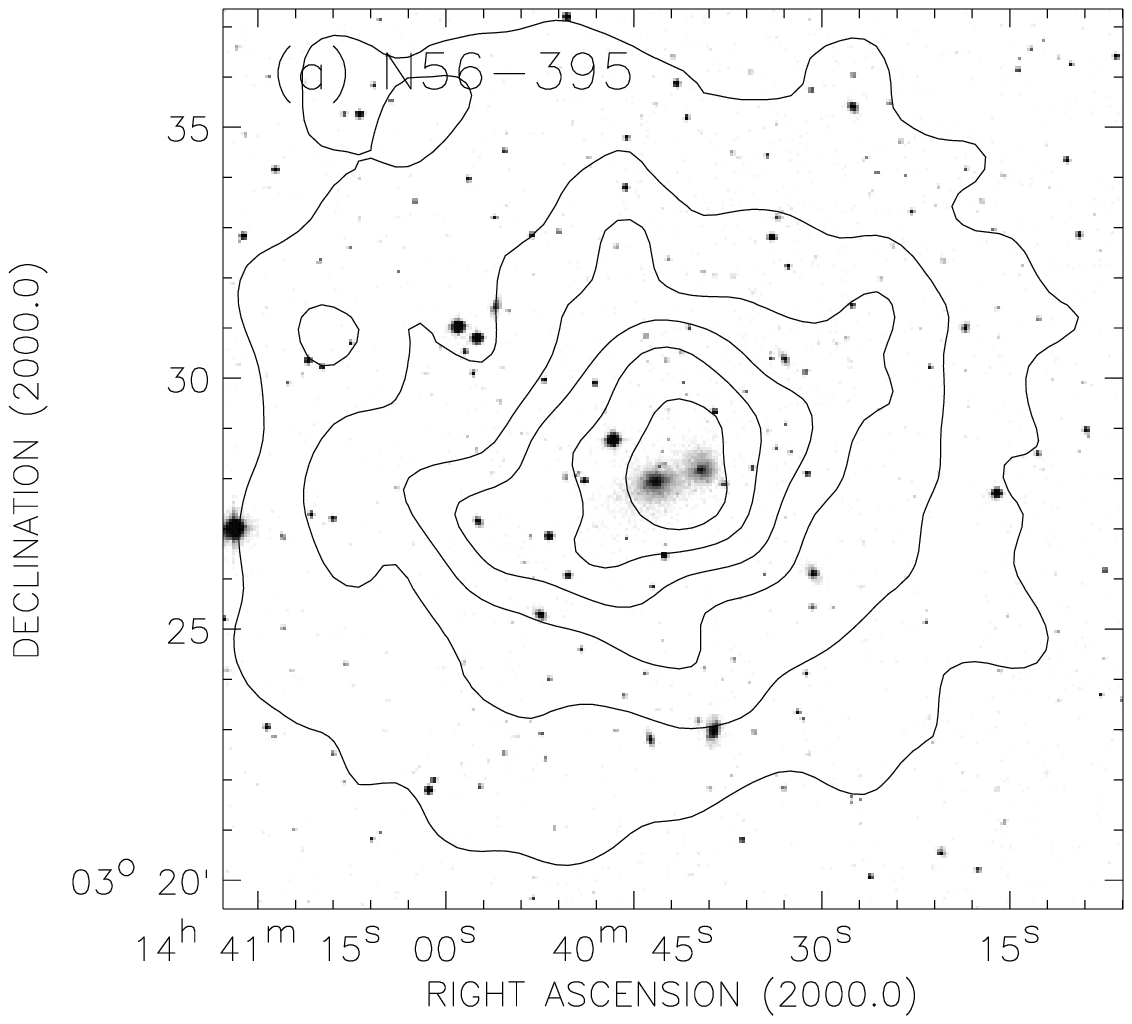,height=2.25truein} \hskip 0.3in
\psfig{figure=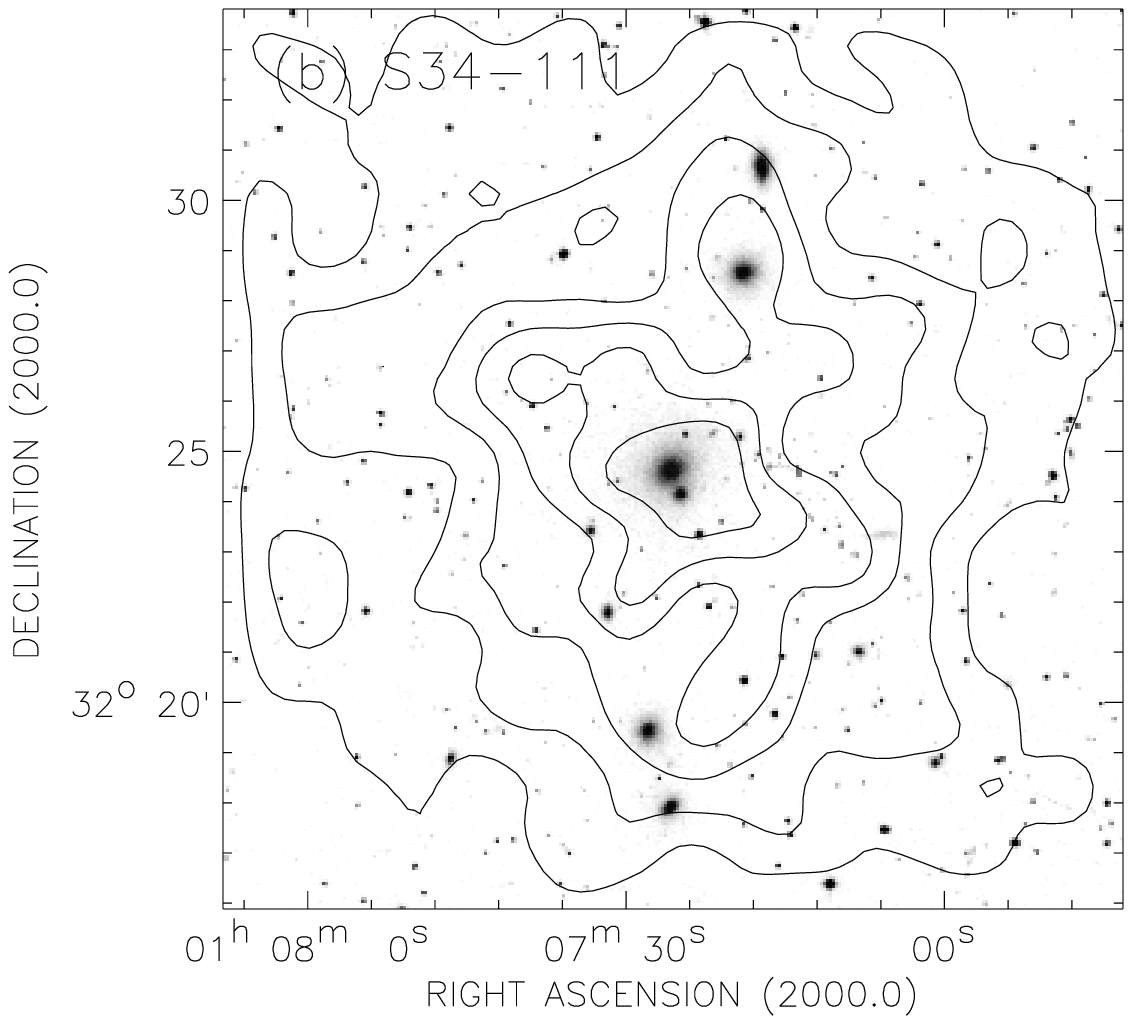,height=2.25truein} \hskip 0.3in
\psfig{figure=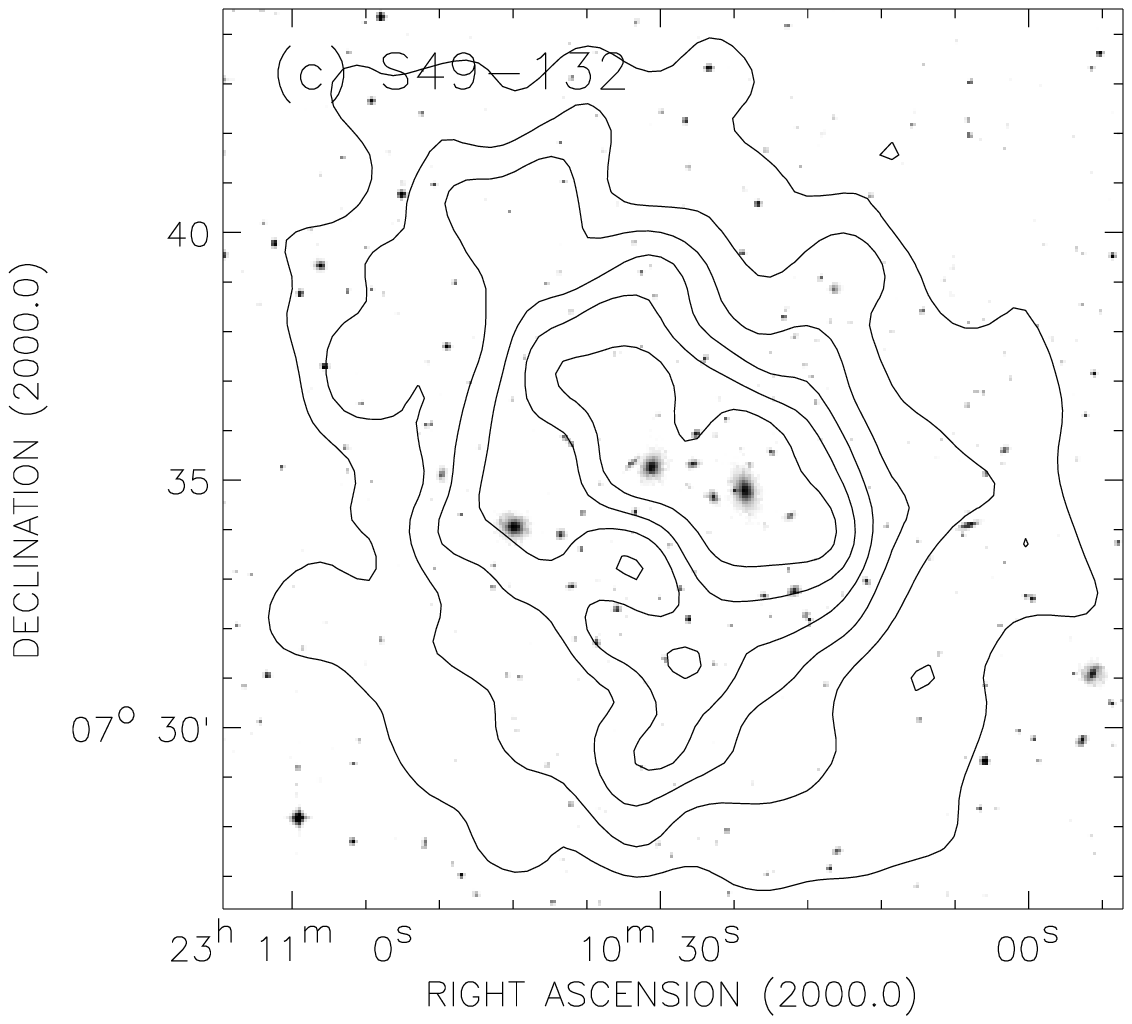,height=2.25truein} 
\hfill
}\vskip 0.3in \centerline{
\hfill 
\psfig{figure=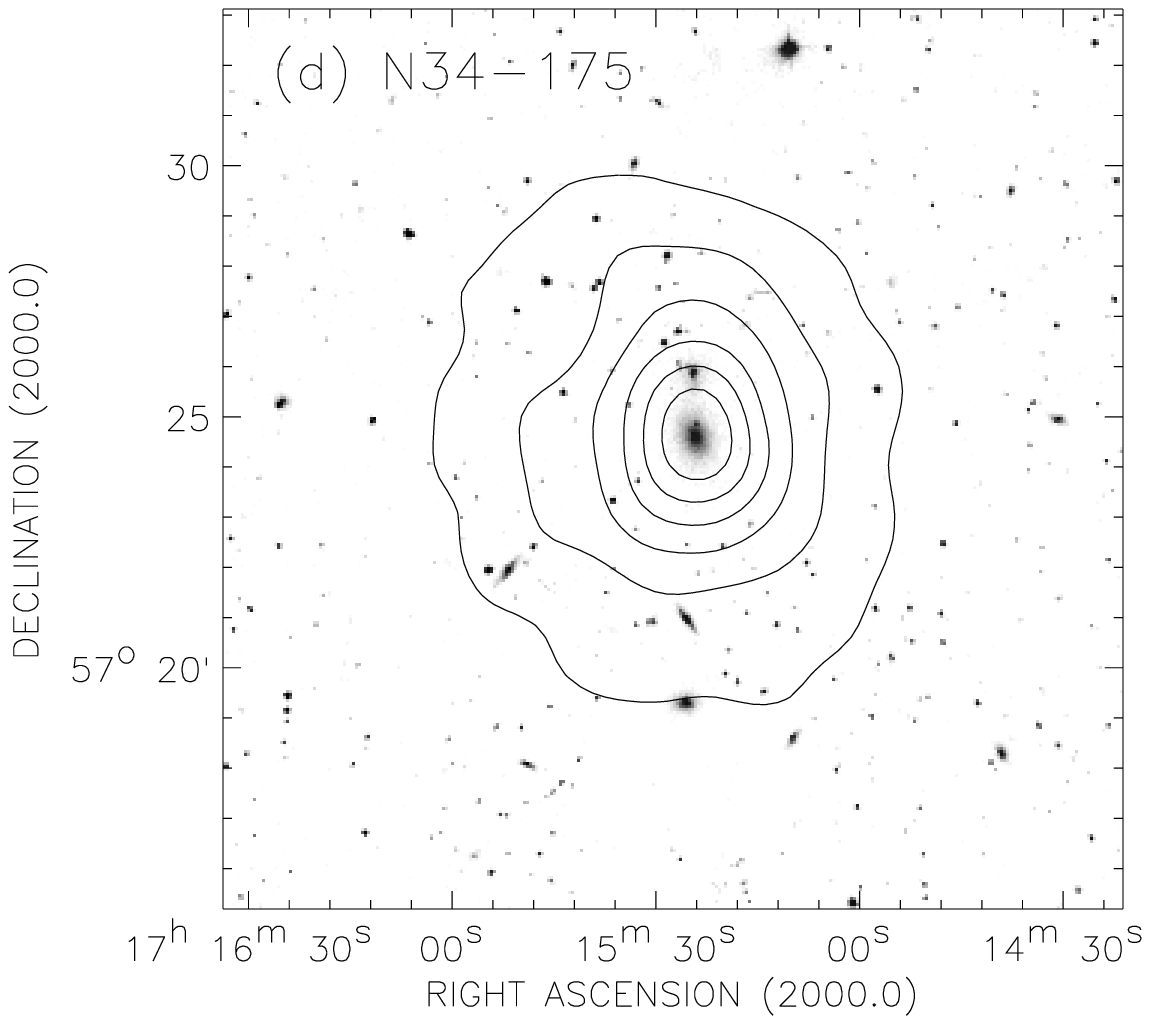,height=2.25truein} \hskip 0.3in
\psfig{figure=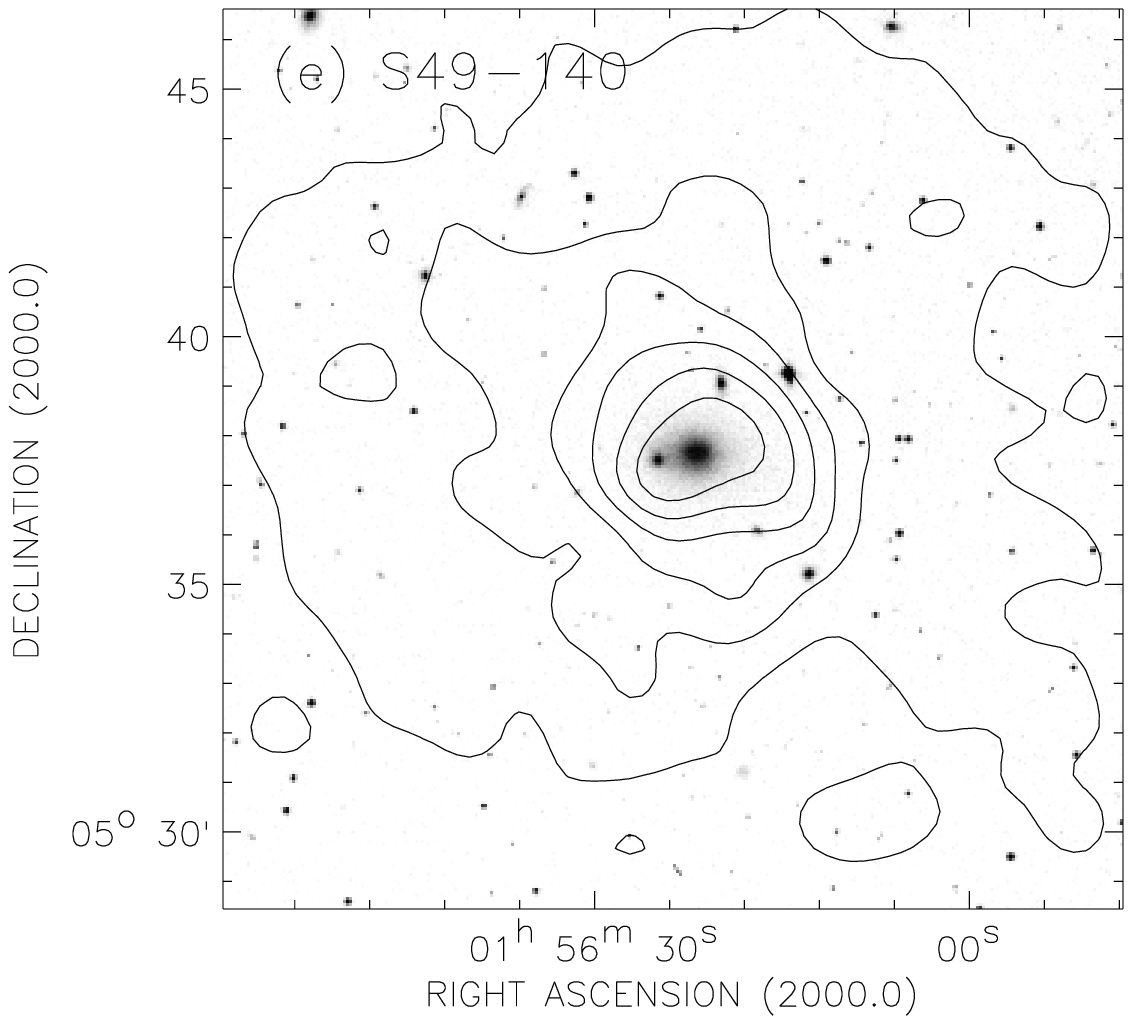,height=2.25truein} 
\hfill
}\caption{X-ray contours from ASCA GIS 2 images overlaid on Palomar
Survey optical images.  Each region shown is 18$'$ across.  The X-ray
images have been smoothed with a Gaussian function with $\sigma=0.6'$.
The contours are linearly spaced.  For N56-395, the minimum and
maximum contours correspond to 4.8$\times 10^{-4}$ and 2.4$\times
10^{-3}$ counts/s/arcmin$^2$; for S34-111, these contours correspond
to 3.4$\times 10^{-4}$ and 1.1$\times 10^{-3}$; for S49-132,
4.8$\times 10^{-4}$ and 1.6$\times 10^{-3}$; for N34-175, 1.1$\times
10^{-3}$ and 6.7$\times 10^{-3}$; for S49-140, 2.2$\times 10^{-4}$ and
1.1$\times 10^{-3}$.}
\end{figure}

\begin{figure}
\centerline{
\hfill 
\hskip 2.5in
\psfig{figure=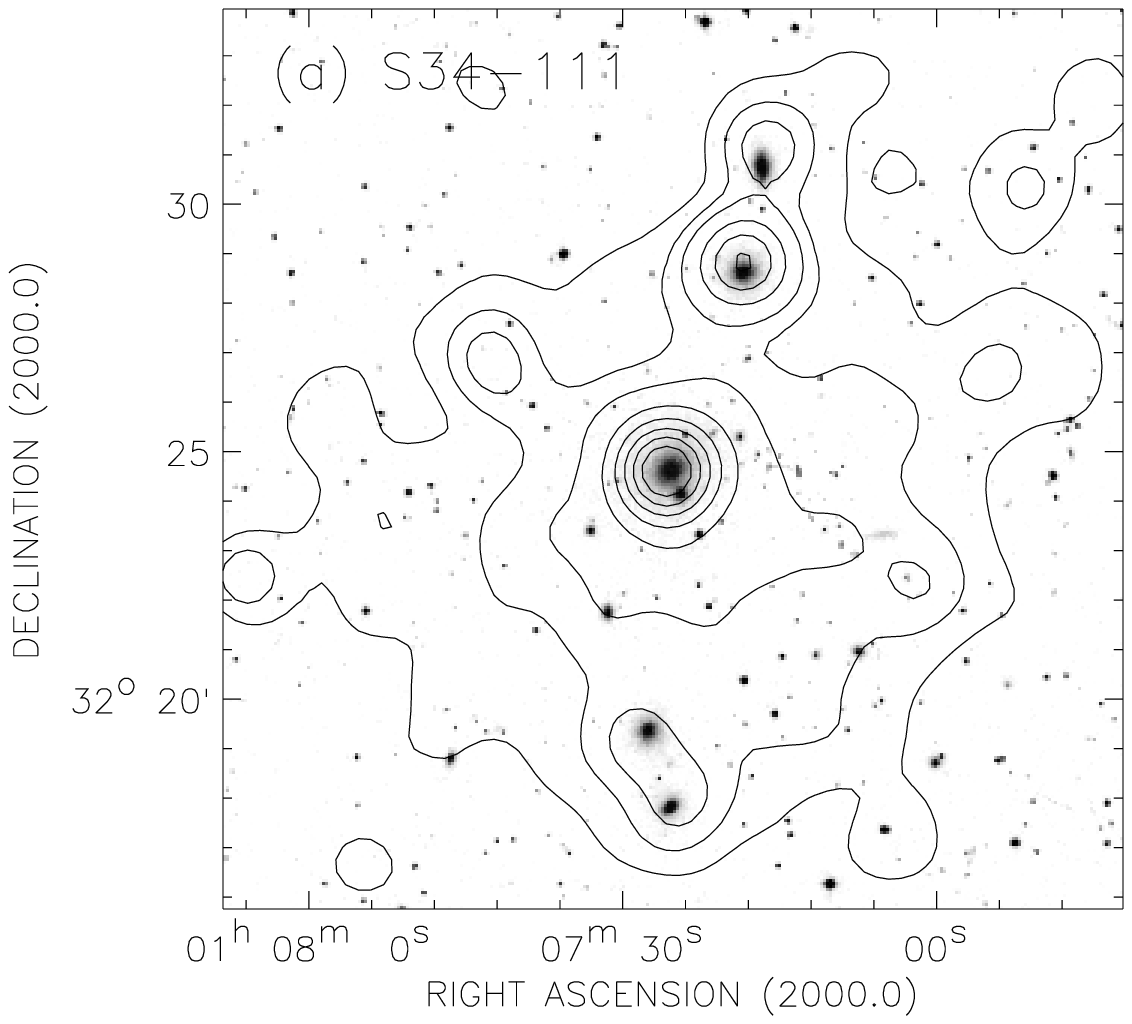,height=2.25truein} \hskip 0.3in
\psfig{figure=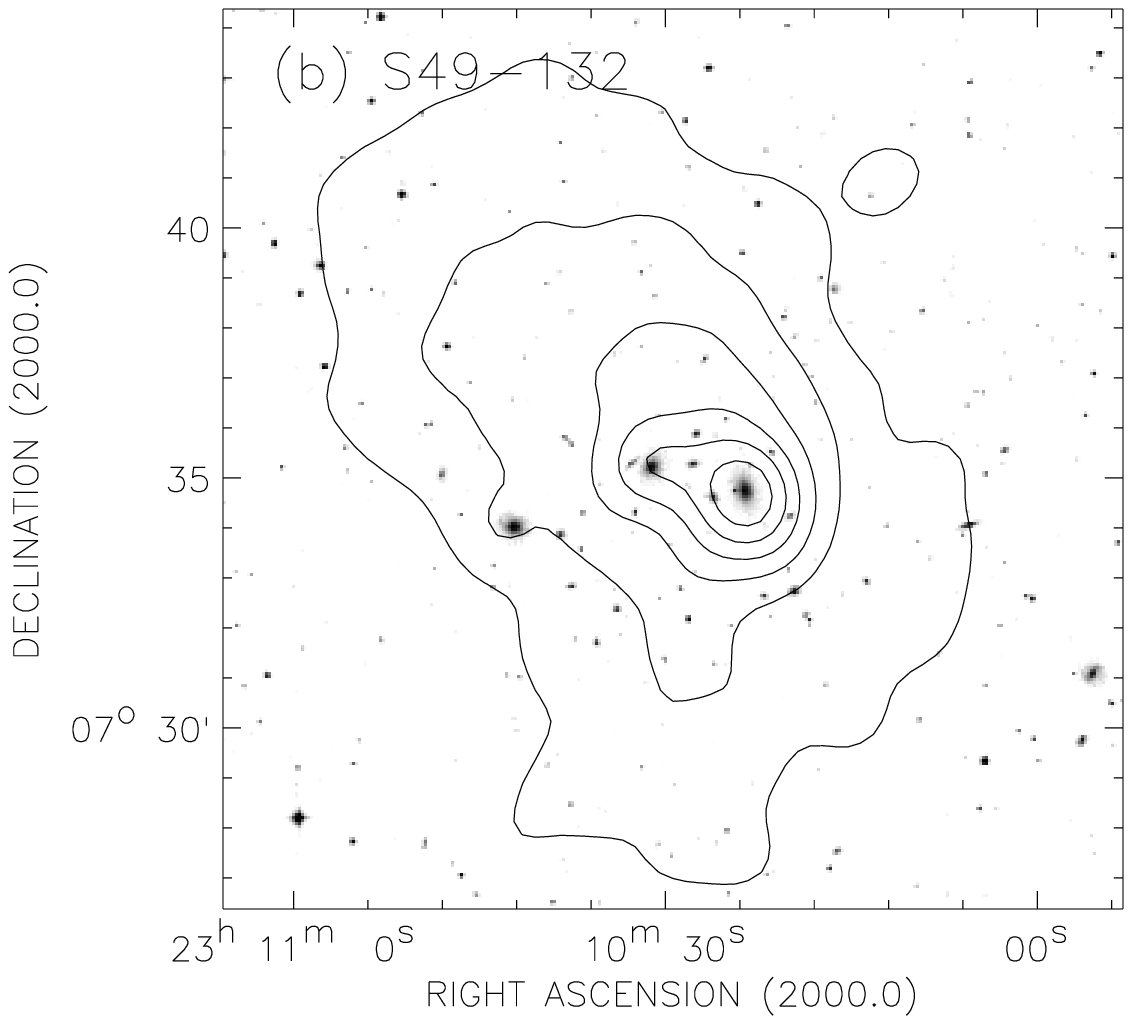,height=2.25truein} 
\hfill
}\vskip 0.3in \centerline{
\hfill
\hskip 2.5in
\psfig{figure=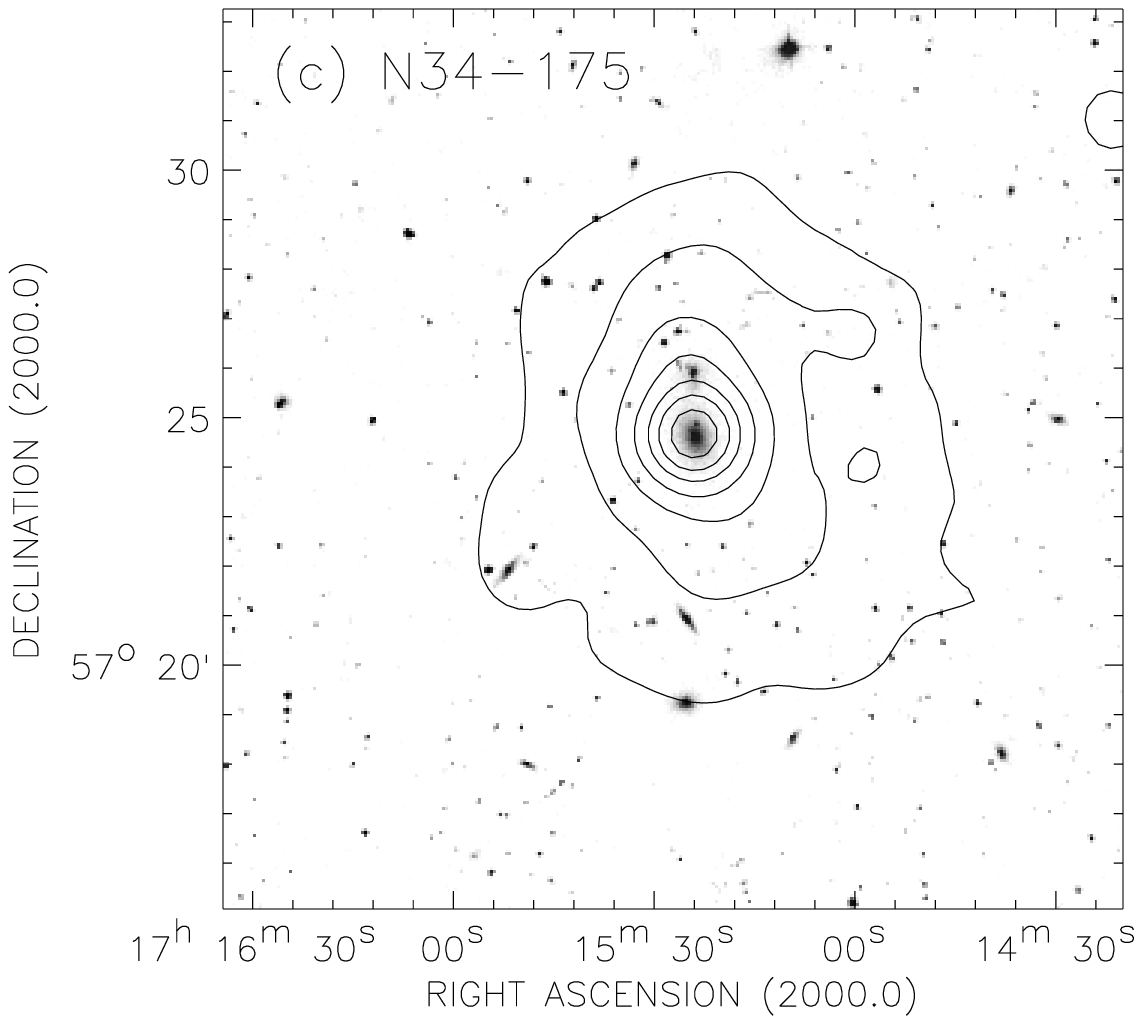,height=2.25truein} \hskip 0.3in
\psfig{figure=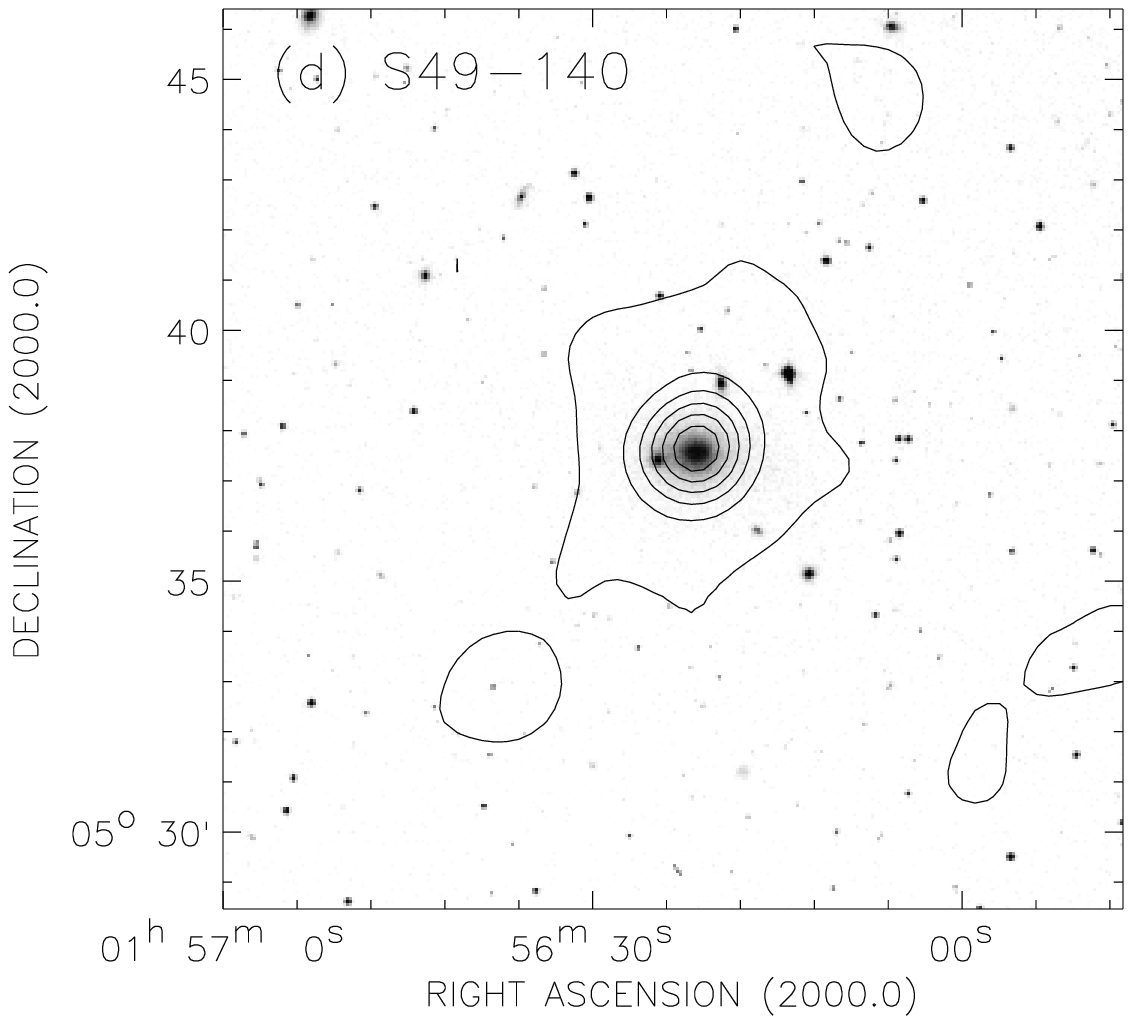,height=2.25truein} 
\hskip 2.5in
\hfill
}\caption{X-ray contours from ROSAT PSPC 0.5-2.0 keV images overlaid
on the same optical images as in Figure 3.  The ROSAT contours for
S34-111 are linearly spaced between 0.001 to 0.007
counts/s/arcmin$^2$, with an additional contour (the second lowest) at
0.0015; for S49-132, between 0.0012 and 0.0069; for N34-175, between
0.002 and 0.045, plus an additional contour (the second lowest) at
0.005; for S49-140, between 0.008 and 0.011.}
\end{figure}

\begin{figure}
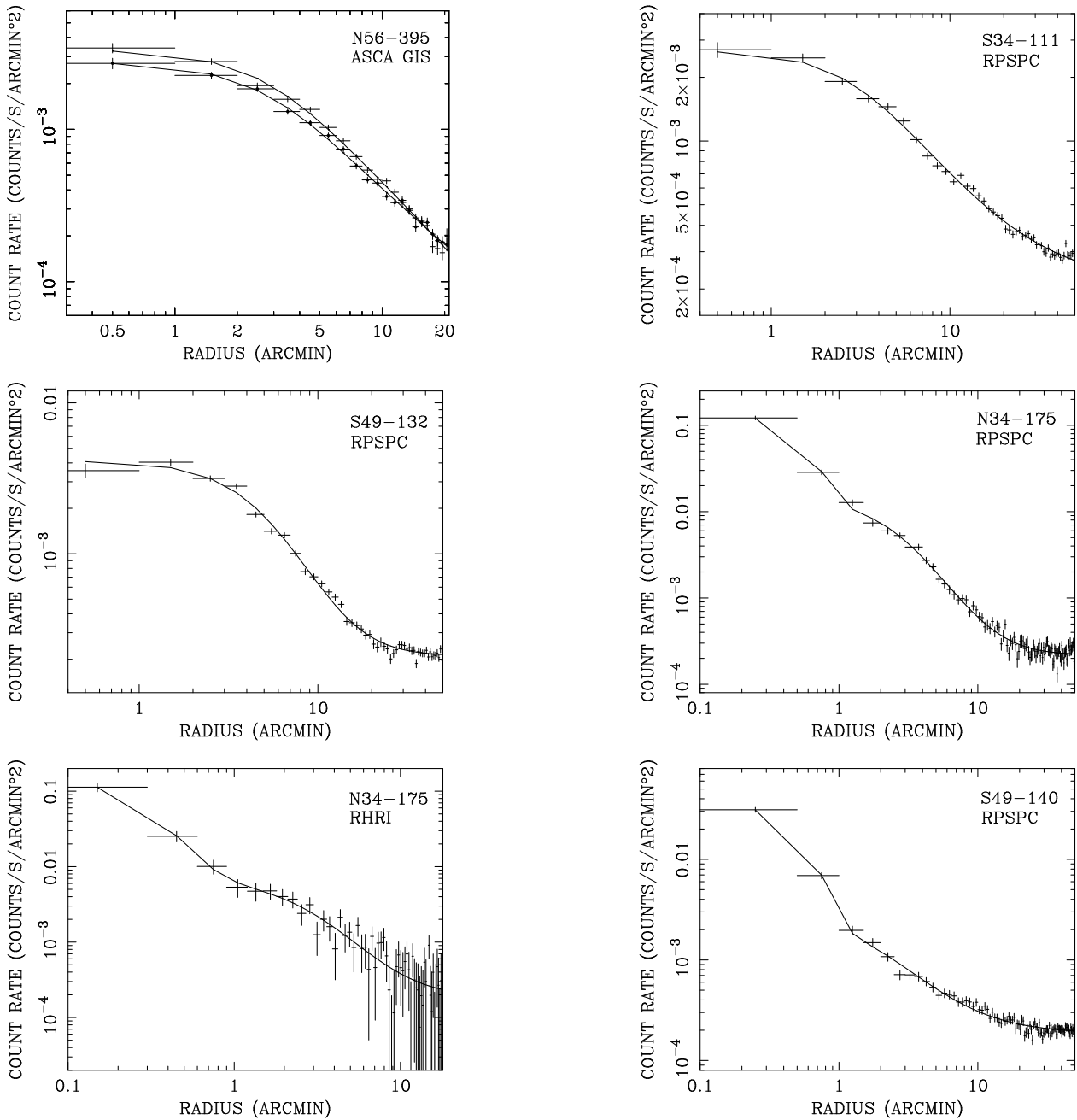

\centerline{
\hskip -0.15in
\psfig{figure=f5a.ps,height=2.2truein,angle=-90} 
\hfill
\psfig{figure=f5b.ps,height=2.2truein,angle=-90} 
}
\vskip 0.1in
\centerline{
\hskip -0.15in
\psfig{figure=f5c.ps,height=2.2truein,angle=-90} 
\hfill
\psfig{figure=f5d.ps,height=2.2truein,angle=-90} 
}
\vskip 0.1in
\centerline{
\hskip -0.15in
\psfig{figure=f5e.ps,height=2.2truein,angle=-90}
\hfill
\psfig{figure=f5f.ps,height=2.2truein,angle=-90}
}\caption{Surface brightness profiles for each group.  The profiles
are from the ROSAT PSPC except for N56-395, which are from the ASCA
GIS2 and GIS3, and for N34-175, for which we show both PSPC and HRI
profiles.  Point sources were either removed from the ROSAT images
before calculating the surface brightness profile (S34-111 and
S49-132), or the central galaxy was modelled as an additional gaussian
component (N34-175 and S49-140).  No point-source subtraction was
carried out for the N56-395 ASCA data.  The best-fit isothermal King
model for the diffuse gas is shown overlaid on the observed surface
brightness profile (see Table 4).}
\end{figure}

\begin{figure}
\centerline{ 
\hfill \psfig{figure=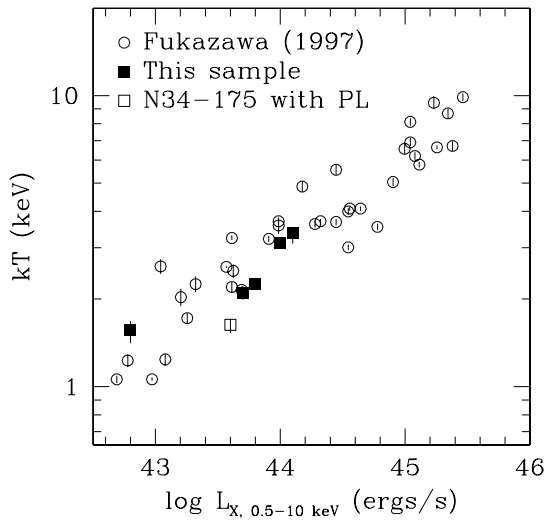,height=3.0truein}
\hfill 
}\caption{X-ray luminosities between 0.5$-$10 keV at 0.5 Mpc and X-ray
measured gas temperatures for clusters and groups in the ASCA sample
of Fukazawa (1997).  Our groups are shown in this plot as the solid
square points.  The open square point represents N34-175 when a
power-law is included in the spectral model (see text).}
\end{figure}

\begin{figure}
\centerline{
\hfill 
\psfig{figure=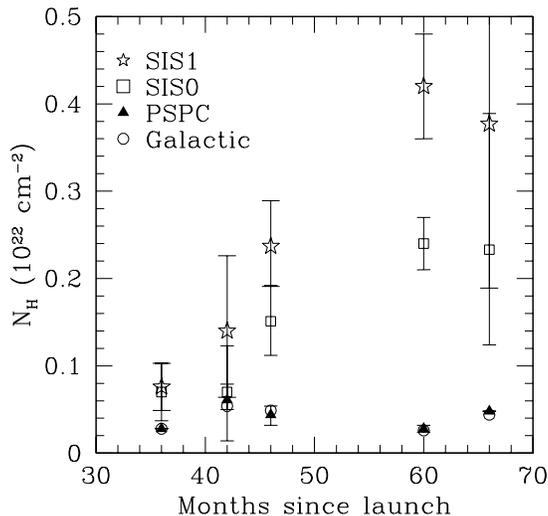,height=3.0truein}
\hfill
}\caption{(Appendix) The fitted BRIGHT mode column densities for SIS0
(square) and SIS1 (star) plotted against observation date measured in
months since the February 1993 ASCA launch along with corresponding
column densities from ROSAT PSPC (triangle) and the radio-measured
Galactic value (circle).  Left to right are: N56-395, S34-111,
S49-132, N34-175, and S49-140.}
\end{figure}

\begin{figure}
\centerline{ 
\hfill \psfig{figure=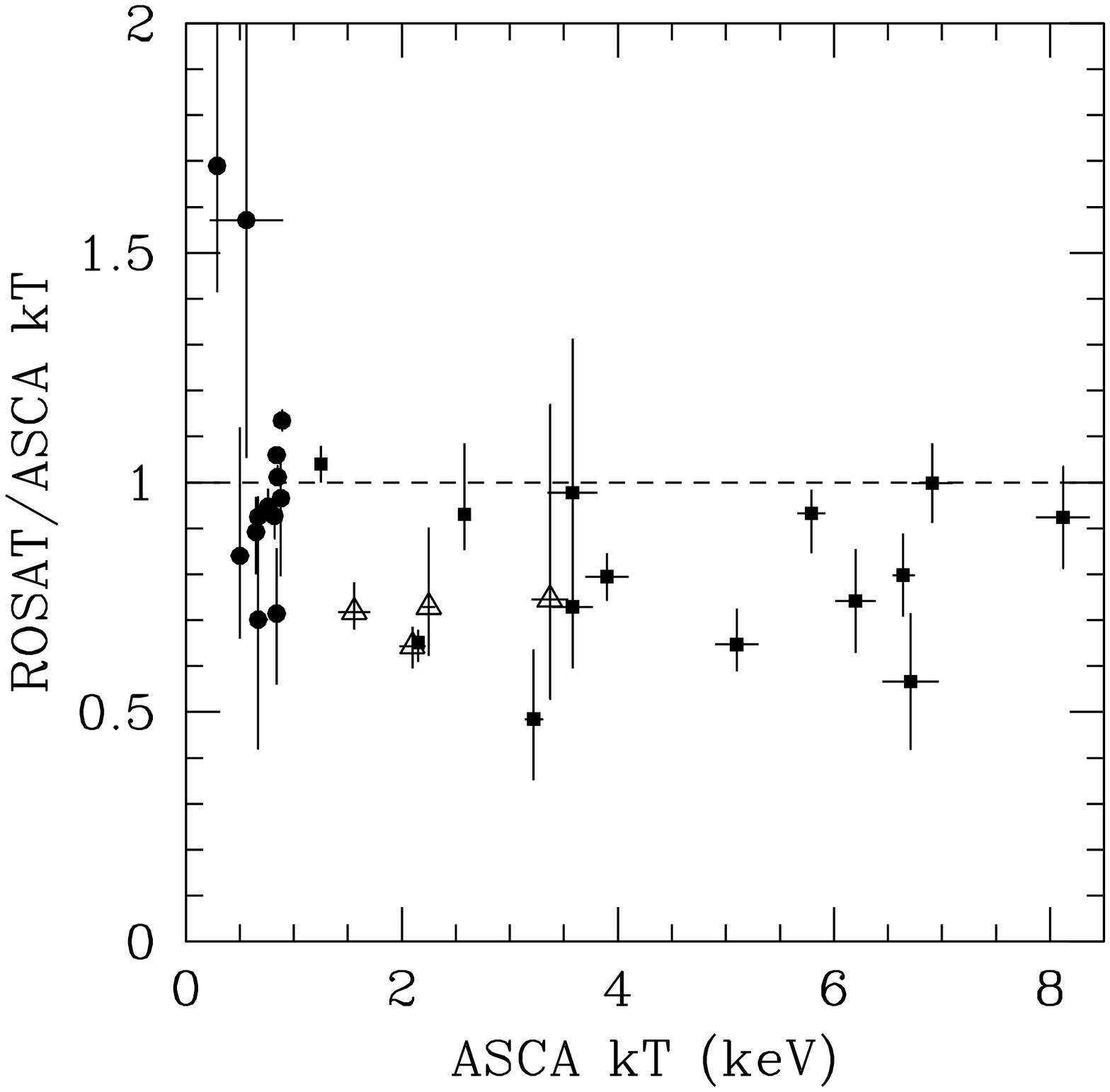,height=3.2true in} \hfill
}\caption{(Appendix) Comparison of temperatures measured by ASCA with
the ratio of ROSAT and ASCA temperatures.  The horizontal errorbars
give the error in the ASCA temperature, the vertical errorbars, the
error in the PSPC temperature only.  The open triangles are the groups
in this study.  The filled circles are for elliptical galaxies, where
the ASCA temperatures are taken from Matsushita (1996) and Matsumoto
et al. (1997), ROSAT temperatures from Davis \& White (1996).  The
filled squares are for clusters: ASCA temperatures are taken from
Fukazawa (1997), and ROSAT temperatures from Bardelli et al. 1996,
Briel \& Henry 1996, David, Jones, \& Forman 1996, Henry, Briel, \&
Nulsen 1993, Jones et al. 1997, Markevitch \& Vikhlinin 1997, Pislar
et al. 1997, Sarazin \& McNamara (1997), and Schindler \& Prieto
1997.}
\end{figure}


\begin{references}
\small{
\baselineskip=0pt\itemsep=0pt
\reference{} Allen, S. W., \& Fabian, A. C. \ 1988, MNRAS, 297, L63
\reference{} Anders, E., \& Grevesse, N. \ 1989, Geo. et Cos. Acta, 53, 197
\reference{} Andersen, V., \& Owen, F. N. \ 1995, AJ, 109, 1582
\reference{} Arnaud, M., Rothenflug, R., Boulade, O. P., Vigroux, L., 
   \& Vangioni-Flam, E. \ 1992, A\&A, 254, 49
\reference{} Bardelli, S., Zucca, E., Malizia, A., Zamorani, G., Scaramella, 
   R., Vettolani, G. \ 1996, A\&A, 305, 435
\reference{} Beers, T. C., Kriessler, J. R., Bird, C. M., \& Huchra, J. P.
   \ 1995, AJ, 109, 874
\reference{} Briel, U. G., \& Henry, J. P. \ 1996, ApJ, 472, 131
\reference{} Burns, J. O., et al. \ 1996, ApJL, 467, L49
\reference{} David, L., Jones, C., \& Forman, W. \ 1996, ApJ, 473, 692
\reference{} David, L., Jones, C., Forman, W., \& Daines, S. \ 1994, ApJ, 
   428, 544
\reference{} Davis, D. S., Mushotzky, R. F., Mulchaey, J. S., Worrall, D. M., 
   Birkinshaw, M., \& Burstein, D. \ 1995, ApJ, 444, 582
\reference{} Davis, D. S., Mulchaey, J. S., Mushotzky, R. F., \& Burstein, D. 
   \ 1996, ApJ, 480,  601
\reference{} Davis, D. S., Mulchaey, J. S., \& Mushotzky, R. F. \ 1998, ApJ, 
   in press
\reference{} Davis, D. S., \& White, R. E. \ 1996, astro-ph/9607052
\reference{} Day, C. S., Arnaud, K. A., Ebisawa, K., Gotthelf, E. V., 
   Ingham, J., Mukai, K., \& White, N. 1995, ABC Guide to ASCA Data
   Reduction, ASCA Guest Observer Facility, NASA/GSFC
\reference{} Dell'Antonio, I. P., Geller, M. J., \& Fabricant, D. G. 
   \ 1994, AJ, 107, 427
\reference{} Dickey, J. M., \& Lockman, F. J. \ 1990, ARAA, 28, 215
\reference{} Doe, S., Ledlow, M. J., Burns, J. O., \& White, R. A. 
   \ 1995, AJ, 110, 46
\reference{} Edge, A. C., \& Stewart, G. C. \ 1991, MNRAS, 252, 414 
\reference{} Edge, A. C., \& Stewart, G. C. \ 1991b, MNRAS, 252, 428
\reference{} Evrard, A. E., Metzler, C. A., \& Navarro, J. F. \ 1996, 
   ApJ, 469, 494
\reference{} Forman, W., Jones, C., \& Tucker,  W. \ 1985, ApJ, 293, 102
\reference{} Fukazawa, Y. \ 1997, PhD Thesis, University of Tokyo 
\reference{} Fukazawa, Y., et al. \ 1996, PASJ, 48, 395
\reference{} Fukazawa, Y., Ohashi, T., Tamura, T., Ikebe, Y., White, R. E.,
   \& Makishima, K. \ 1996b, UV and X-ray Spectroscopy of Astrophysical and
   Laboratory Plasmas (ed. Yamashita, K., \& Watanabe, T.), Tokyo: Universal
   Academy Press, 383
\reference{} Gregory, P. C., \& Condon, J. J. \ 1991, ApJS, 75, 1011
\reference{} Henry, J. P., et al. \ 1995, ApJ, 449,422
\reference{} Henry, J. P., Briel, U. G., \& Nulsen, P. E. J. \ 1993, A\&A,
   271, 413
\reference{} Isobe, K., Tawara, Y., Yamashita, K., Kunieda, H., \& Watanabe,
   M. \ 1996, X-ray Imaging and Spectroscopy of Cosmic Hot Plasmas (ed. 
   Makino, F., \& Mitsuda, K.), Tokyo: Universal Academy Press, 175
\reference{} Jones, C., Stern, C., Forman, W., Breen, J., David, L., 
   Tucker, W., \& Franx, M. \ 1997, ApJ, 482, 143
\reference{} Ledlow, M. J., Loken, C., Burns, J. O., Hill, J. M., 
   \& White, R. A. \ 1996, AJ, 112, 388
\reference{} Loewenstein, M., \& White, R. E. III \ 1998, ApJ, submitted
\reference{} Mahdavi, A., Bohringer, H., Geller, M. J., \& Ramella, M. 
   \ 1997, ApJ, 483, 68
\reference{} Markevitch, M., \& Vikhlinin, A. \ 1997, ApJ, 474, 84
\reference{} Matsumoto, H., Koyama, K. Awaki, H., Tsuru, T., Loewenstein, 
   M., \&  Matsushita, K. \ 1997, ApJ, 482, 133
\reference{} Matsushita, K. \ 1996, PhD Thesis, University of Tokyo
\reference{} McElroy, D. B. \ 1995, ApJS, 100, 105
\reference{} Mulchaey, J. S., Davis, D. S., Mushotzky, R. F., \& Burstein, D.
   \ 1996, ApJ, 456, 80
\reference{} Mulchaey, J. S., \& Zabludoff, A. I. \ 1998, ApJ, 496, 73
\reference{} Mushotzky, R. F. \ 1984, Phys. Scr., T7, 157
\reference{} Mushotzky, R. F., Loewenstein, M., Arnaud, K. A., Tamura, T., 
   Fukazawa, Y., Matsushita, K., Kikuchi, K., \& Hatsukade, I. \ 1996,  
   ApJ, 466, 686
\reference{} Mushotzky, R. F., \& Scharf, C. A. \ 1997, ApJL, 482, 13
\reference{} Pedersen, K., Yoshii, Y., Sommer-Larsen, J. \ 1997, ApJL,
   485, 17
\reference{} Persic, M., Salucci, P., \& Stel, F. \ 1996, MNRAS, 281, 27
\reference{} Pildis, R., A., Bregman, F. N., \& Evrard, A. E. \ 1995, 
   ApJ, 443, 514
\reference{} Pislar, V., Durret, F., Gerbal, D., Limaneto, G. B., 
   \& Slezak, E. \ 1997, A\&A, 322, 53
\reference{} Ponman, T. J., \& Bertram, D. \ 1993, Nature, 363, 51
\reference{} Ponman, T. J., Bourner, P. D. J., Ebeling, H., \& Bohringer, 
   H. \ 1996, MNRAS, 283, 690
\reference{} Price, R., Duric, N., Burns, J. O., \& Newberry, M. V. \ 1991,
   AJ, 102, 14
\reference{} Quintana, H., \& Melnick, J. \ 1982, AJ, 87, 972
\reference{} Raymond, J. C., \& Smith, B. W. 1977 \apjs, 35, 419
\reference{} Renzini, A. \ 1997, ApJ, 488, 35
\reference{} Sakima, Y., Tawara, Y., \& Yamashita, K. \ 1995, New Horizon
   of X-ray Astronomy (ed. Makino, F., \& Ohashi, T.), 557
\reference{} Sarazin, C., L. \& McNamara, B. R. \ 1997, ApJ, 480, 203
\reference{} Schindler, S. \ 1996, A\&A, 305, 858
\reference{} Schindler, S., \& Prieto, M. A. \ 1997, A\&A, 327, 37
\reference{} Snowden, S. L., McCammon, D., Borrows, D. N., \& Mendenhall, 
   J. A. \ 1994, ApJ, 424, 714
\reference{} Tanaka, Y., Inoue, H., \& Holt, S. S.  1994 \pasj, 46, L37
\reference{} Trussoni, E., Massaglia, S., Ferrari, R., Fanti, R., Feretti, 
   L., Parma, P., \& Brinkmann, W. \ 1997, \aap, 327, 27
\reference{} White, D. A., Jones, C. , \& Forman, W. \ 1997, MNRAS, 292, 
   419
\reference{} White, R.A., Bliton, M., Bhavsar, S., Bornmann, P., Burns,
   J.O., Ledlow, M., Loken, C. \ 1998, AJ, to be submitted
\reference{} Zabludoff, A. I., \& Mulchaey, J. S. \ 1998, ApJ, 496, 39
}
\end{references}
\end{document}